\documentstyle{article}
\begin{document}
\pagestyle{empty}
\Large
\centerline{\bf{The Energies of All Hadrons(Including All Known
Resonances)}}
\centerline{\bf{and the Energies of the Excited States of Quarks}}
\vskip .3in
\noindent
\normalsize
\centerline{M\'{a}rio Everaldo de Souza,}
\newline
\noindent
\centerline{Departamento de F\'{\i}sica - CCET, Universidade Federal de
Sergipe,}
\newline
\noindent
\centerline{Campus Universit\'{a}rio, 49000 Aracaju, Sergipe, Brazil}
\vskip .6in
\noindent
\centerline{ABSTRACT}
\vskip .2in
\par By means of a general classification of the different kinds of matter
which were formed along the universal expansion this paper shows that the
forces of nature form a chain from the world of the subatomic particles to the
large bodies of the universe, the galaxies.
\par It is shown that matter has a generalized structured state characterized
by the existence of some degree of order and by some kind of Lennard-Jones
potential. The existence of such a state at the level of quarks and galaxies
suggests that nature has two more fundamental forces, a superstrong interaction
which acts among quarks and prequarks, and a superweak interaction which
acts among galaxies.
\par The paper also gives a physical explanation for quark
confinement. The energies of hadrons, including all known resonances, are
calculated in a simple manner. One may predict the energies of all the other
hadrons to be found experimentally. The excited states of quarks are also
calculated. According to the work there is no need for the Higgs boson since
the masses of hadrons and quarks are generated by the superstrong and strong
interactions.
\newpage
\noindent
I. INTRODUCTION
\vskip .15in
\par It has been proven beyond any doubt that the universe is expanding
${(1,2,3,4)}$. Recent data of several investigators show that  galaxies form
gigantic structures in space. De Lapparent et al.${(5)}$(also, other papers by
the same authors) have shown that they form bubbles which contain huge voids of
many megaparsecs of diameter. Broadhurst et al.${(6)}$ probed deeper regions
of the universe with two pencil beams and showed that there are (bubble) walls
up to a distance of about 2.5 billion light-years from our galaxy.  Even more
disturbing is the apparent regularity of the walls with a period of about
$130h{-1}$Mpc. There is also an agglomeration of galaxies forming a thicker
wall, called the great wall. It has also been reported${(7,8,9)}$ (also, other
confirming papers) that there is a large-scale coherent flow towards a region
named the Great Attractor. All this data show that  galaxies form a medium in
which we already observe some interactions.  Kurki-Suonio et al.${(10)}$
the medium formed by them is rather a liquid than a crystalline solid. We may
call this medium the `galactic liquid'.
\par The facts shown above have also been corroborated by the data of the APM
survey${(11)}$ which was based on a sample of more than two million galaxies.
This survey measured the galaxy correlation function $w(\theta)$.
More evidence towards the same conclusion was provided by the `counts in cells'
of the QDOT survey$ {(12)}$ and more recently by the redshift(APM)
survey$ {(13)}$ and by the power spectrum inferred from the CfA survey$
{(14)}$and from that inferred from the Southern Sky Redshift Survey$ {(15)}$.
The  data from these surveys show that galaxies are not randomly distributed in
space on large scales and put a shadow over the cold dark matter
model${(16,17)}$. There have been attempts to keep the cold dark matter
hypothesis but only at the cost of many {\it{ad hoc}} adjustments.  There are
even proposals of having more than one type of dark matter$ {(18)}$.
\par Another puzzle of Nature is the existence of strange objects, such as
quasars, BL Lacertae and Seyfert galaxies, and the fact of having spiral
galaxies with their mysterial arms. The evolution of galaxies does not fit
either in the general theoretical framework.
\par At the other end of the distance scale, in the fermi region, it appears
now that the quark is not elementary after all. This can be implied just
from their number which, now, stands at 18. Particle physics theorists
have already begun making models addressing this compositeness$
{(19)}$.
\vskip .3in
\noindent
II. GENERAL CLASSIFICATION OF MATTER
\vskip .2in
\par Science has utilized specific empirical classifications of matter
which have revealed hidden laws and symmetries.  Two of the most known
classifications are the Periodic Table of the Elements and Gell-Mann's
classification of particles(which paved the way towards the quark model).
\par Let us go on the footsteps of Mendeleev and let us attempt to achieve a
general classification of matter, including all kinds of matter formed along
the universal expansion, and by doing so we may find the links between the
elementary particles and the large bodies of the universe.
\par It is well known that the different kinds of matter of nature appeared
at different epochs of the universal expansion, and that, they are imprints of
the different sizes of the universe along the expansion.
Taking a closer look at the different kinds of matter  we may classify them
as belonging to two distinct general states. One state is characterized by a
single unity with  angular momentum, and we may call it, the single state.
The angular momentum may either be the intrinsic angular momentum, spin, or the
orbital angular momentum. The other state is characterized by some degree of
correlation among the interacting particles and may be called
the structured state. The angular momentum may(or may not) be present in this
state.  In the single states we find the fundamental unities of matter that
make the structured states. The different kinds of fundamental matter are the
building blocks of everything, {\it{stepwise}}.
In what follows we will not talk about the weak force since it
does not form any stable matter and is rather related to instability in matter.
As discussed in this work it appears that along the universal expansion
nature made different building blocks which filled the space.  The weak force
did not form any building block and is out of the initial discussion.
As is well known this force is special in many other ways.  For example, it
violates parity in many decays and it has no ``effective potential''(or static
potential) as the other interactions do.  Besides, the weak force is known
to be left-handed, that is, particles experience this force only when their
spin direction is anti-aligned with their momentum.  Right-handed particles
appear to experience no weak interaction, although, if they have electric
charge, they may still interact electromagnetically. Later on we will include
the weak force into the discussion. The single state is made by only one kind
of fundamental force. In the structured state one always finds two types of
fundamental forces, i.e., this state is a link between two single states. Due
to the interactions among the bodies(belonging to a particular single state)
one  expects other kinds of forces in the structured state. In this fashion
we can form a chain from the quarks to the galactic superstructures and
extrapolate at the two ends towards the constituents of quarks and towards the
whole universe.
\par The kinds of matter belonging to the single states are the nucleons,
the atom, the galaxies, etc. The `et cetera' will become clearer later on
in this article. In the structured state one finds the quarks, the nuclei, the
gasses, liquids and solids, and  the galactic liquid. Let us, for example,
examine the sequence nucleon-nucleus-atom. As is well known up to now
nucleon is made out of quarks and held together by means of the strong
force. The atom is made out of the nucleus and the electron(we will talk about
the electron later), and is held together by means of the
electromagnetic force. The nucleus, which is in the middle of the sequence, is
held together by the strong force(attraction among nucleons) and by the
electromagnetic force(repulsion among protons). In other
words, we may say that the nucleus is a compromise between these two forces.
Let us, now, turn to the sequence atom-(gas,liquid,solid)-galaxy. The gasses,
liquids and solids are also formed by two forces, namely, the electromagnetic
and the gravitational forces. Because the gravitational force is $10{39}$
weaker than the electromagnetic force the polarization in gasses, liquids and
solids is achieved by the sole action of the electromagnetic force because
it has two signs. But it is well known that large masses of gasses, liquids
and solids are unstable configurations of matter in the absence of gravity.
Therefore, they are formed by the electromagnetic and gravitational forces.
The large amounts of hydrogen {\it{gas}} at some time in the history of the
universe gave origin to galaxies which are the biggest individual unities of
creation. We arrive again at a single fundamental force that holds a galaxy
together, which is the gravitational force. There is always the same pattern:
one goes from one fundamental force which exists in a single unity(nucleon,
atom, galaxy) to two fundamental forces which coexist in a medium. The
interactions in the medium form a new unity in which the action of another
fundamental force appears. We are not talking any more about the previous unity
which exists inside the new unity(such as the nucleons in the nucleus of an
atom).
\par By placing all kinds of matter together in a table in the order of the
{\it{universal expansion}}  we can construct the two tables below,
one for the states and another for the fundamental forces.
\par In order to make the atom we need the electron besides the nucleus.
Therefore, just the clumping of nucleons is not enough in this case.
Let us just borrow the electron for now.  Therefore, it looks like that the
electron belongs to a separate class and is an elementary particle.
The above considerations may be summarized by the following: {\it{the different
kinds of building blocks of the Universe(at different times of the expansion)
are intimately related to the idea of filling space.}}  That is, depending on
its size, the Universe is filled with different unities.
\vskip .3in
\noindent
III. THE NUMBER OF FUNDAMENTAL FORCES OF NATURE
\vskip .2in
\par In order to keep the same pattern, which should be related to an
underlying symmetry, the tables reveal that there should be another force,
other  than the strong force, holding the quarks together, and that this force
should hold together the prequarks. Let us name it the superstrong force.
Also, for the `galactic liquid', there must be another fundamental force at
play. Because it must be much weaker than the gravitational force(otherwise,
it would already have been found on Earth) we expect it to be a very weak
force.  Let us call it the superweak force.
\par Summing up all fundamental forces  we arrive at {\it{six forces for
nature: the superstrong, the strong, the electromagnetic, the gravitational,
the superweak and the weak  forces}}.  We will see later on that they are
interrelated.
\noindent
IV. THE STRUCTURED STATE
\vskip .2in
\par The structured state is made by opposing forces, i.e., it represents a
compromise between an attractive force and a repulsive force.  Ordinary
matter(gasses, liquids and solids) is formed by the polarization of the
electromagnetic force.  Polarization is also present in nuclear matter which
may be described either in terms of the Seyler-Blanchard interaction or
according to the Skirme interaction. Both  give a type of Van der Waals
equation of state${(20,21)}$.  It is not by chance, then, that the liquid
drop model provides quite some satisfactory results in nuclear physics. In
order to keep the same pattern we should expect to have a sort of compromise
between the superstrong force and the strong force. This compromise forms the
quark. As we will show later the superstrong interaction is mediated by three
bosons. By analogy with the polarization of the electromagnetic force among
molecules, we expect to have some sort of polarization among quarks also. Now,
we can understand why lattice QCD yields many satisfactory results.
\par In order to have the `galactic liquid' it is also necessary to have
some sort of polarization.  This means that we need dipoles and because the
gravitational force is always attractive(and thus, can not be the source of
such dipoles), the superweak force must be repulsive during the universal
expansion.  This is consistent with the idea of the expansion itself.  That is,
the universal expansion must be caused by this repulsive force.
\par The bodies which form any structured state exhibit some degree of
correlation among them. This degree of correlation is shown by the correlation
function which, in turn, is related to the interacting net potential energy
among the particles.  The potential energy has three general features: i) it
has a minimum which is related to the mean equilibrium positions of the
interacting particles; ii) it tends to zero as the separation among the
particles tends to infinity; and iii) it becomes repulsive at close distances.
A good illustration of the general features of such a potential energy is the
molecular potential(Fig.1). Because of our ignorance in treating the many-body
problem, it has become quite commom the use of the so-called semi-empirical
interatomic potential energies. Mathematically, there are a few of them. The
most commonly used is the Lennard-Jones potential energy${(22)}$ which has the
general form
\begin{eqnarray} V(r) &=& \frac{\lambda_n}{r n} - \frac{\lambda_m}{r m}.
\end{eqnarray}
\noindent
Later on we will come back to this point. The potential energy of the `galactic
liquid' must also have the same general features.  Therefore, it is very
important to determine the mean equilibrium position of the galactic
superstructures.
\par As is well known the general motion of the particles of a liquid is quite
complex and that is exactly what we are dealing with in the case of the
galactic superstructures.
\par The aggregation of matter into larger and larger structures is possible
because of the higher orders of interaction of polarization, such as
dipole-dipole, dipole-quadrupole, etc. For example, a molecule is formed by
the polarization of their atoms. But there is also a polarization between
two molecules. This also takes place with the constituents of the nucleus.
There is a polarization between two protons and there is also a polarization
between two alpha particles. As discussed above we expect to have a
polarization in the formation of a quark and also in the interaction among
quarks.
\vskip .3in
\noindent
V. PRELIMINARY IDEAS ON PREQUARKS
\vskip .2in
\par The classification of matter achieved above implies that quarks are
formed of prequarks. Let us develop some preliminary ideas which may help us
towards the understanding of the superstrong interaction. Presumably, just as
quarks do, prequarks are also supposed to be permanently confined inside
baryons.
\par Actually, the composition of quarks is an old idea, although it has been
proposed on different grounds${(23)}$. A major distinction is that in this
work leptons are supposed to be elementary particles. This is actually
consistent with the smallness of the electron mass which is already too small
for a particle with a very small radius$ {(24)}$.
\par In order to distinguish the model proposed in this paper from other models
of the literature we will name these prequarks with a different name.   We may
call them {\it{primons}}, a word derived from the latin word {\it{primus}}
which means first.
\par From the above considerations quarks are composed of primons. Also, we
saw above that there must be a sort of polarization in the formation of a
quark.  Since a baryon is composed of three quarks, it is reasonable to
consider that a quark is composed of two primons which are polarized by the
exchange of some kind of charge which is carried on by the corresponding boson.
Let us name the exchanged particle the mixon(as we will see below the quark
colors will come from the mixing of supercolors).
\par In order to reproduce the spectrum of 18 quarks(6 quarks in 3 color
states) we need 12 primons(4 primons in 3 supercolor states).  Therefore, we
have 4 triplets. As to the charge, one triplet has charge (5/6)e and any other
triplet has charge (-1/6)e.
\par Let us assume that the exchanged particle is a boson. Since quarks have
spin 1/2, the spin of each primon has  to be equal to 1/4. In this case the
boson has spin zero. At this stage we may introduce a postulate concerning the
unit of quantization. As is well known the unit of quantization is $\hbar$.
This unit of measure is arbitrary and was taken as such in order to have an
agreement between experimental data and theory at the level of atomic physics.
Later on this unit of measure was applied to elementary particles and up to
the level of quarks it still holds.  However, at the level of the quarks
constituents it may not hold anymore.  We may postulate that at the level of
primons the unit of quantization is $\bar{\hbar}=\hbar/2$. In this way primons
are also fermions with a spin given by
\begin{eqnarray} s&=&\left(\frac{1}{2}\right)\bar{\hbar}.
\end{eqnarray}
\noindent
As we will see below, this is consistent with the required properties which are
needed for forming quarks out of primons.
\par Let us try to arrive at a possible equation for ``free'' primons beginning
with Dirac's equation,
\begin{eqnarray} \left(i{\hbar}{\alpha}_{\mu}\frac{\partial}{\partial{x}
{\mu}}
- {\beta}mc\right)\Psi &=& 0 \end{eqnarray}
\noindent
where  $\mu=0,1,2,3$. If, now, we divide this equation by 2, and make the
substitutions ${\hbar}/2=\bar{\hbar}$ and ${\bar{\beta}}/2=\beta$, we obtain,
of course, another Dirac equation.  By imposing that each component
$\Psi_{\sigma}$ of $\Psi$ must satisfy Klein-Gordon equation we obtain the
following algebra, which is slightly different from Dirac's,
\begin{eqnarray*}\alpha_{i}\alpha_{k} + \alpha_{k}\alpha_{i} &=& 2\delta_{ik}
\nonumber \\
                \alpha_{i}\bar{\beta} + \bar{\beta}\alpha_{i} &=& 0\nonumber \\
                {\alpha_{i}} {2} &=& 1 \nonumber \\
                {\bar{\beta}} {2} &=& \frac{1}{4}.
\end{eqnarray*}
\noindent
Of course, $\alpha_{i}$ are the same as in Dirac's equation, but now
$\bar{\beta}={\beta}/2$.
\par The new Dirac equation becomes
\begin{eqnarray} \left(i\bar{\hbar}\alpha_{\mu}\frac{\partial}{\partial{x}
{\mu}} - {\bar{\beta}}mc\right)\Psi &=&0 \end{eqnarray}
\noindent
where $\bar{\hbar}$ is a new unit of quantization and $\bar{\beta}$ is Dirac's
${\beta}/2$. Of course, this equation is also Lorentz covariant.
\par In terms of the Hamiltonian, the new Dirac equation is
\begin{eqnarray} \left(-ic\bar{\hbar}{\bf{\alpha}.{\nabla}} +
\bar{\beta}mc {2} + \Phi\right)\Psi &=&
i\bar{\hbar}\frac{\partial\Psi}{\partial{t}}\end{eqnarray}
\noindent
where $\Phi$ is the field by means of which primons interact.
\par Since a quark has spin equal to 1/2, only primons with parallel(or
antiparallel) spins form bound states(quarks). This means that the spin wave
function of a bound state(quark) is symmetrical and, because the total wave
function is antisymmetrical, the rest of the wave function(which includes the
superflavor, supercolor and spatial parts) has to be antisymmetrical.
\par The superstrong interaction should be such that only primons with
different quantum numbers form bound states, that is, quarks. Taking into
account the above considerations on spin and charge, we have the following
table for primons(Table 3).  With this table we are able to form all quarks
as shown in Table 4. The colors are formed from the mixing of the
supercolors as shown in Table 5.
\par Therefore,  a prequark, $p_{ij}$, must transform with
\begin{itemize}
\item superflavor index: $i=1,2,3,4$
\item supercolor index: $j={\alpha},{\beta},{\gamma}$.
\end{itemize}

These indices must transform respectively under $
SU(4)|_{superflavor}$ and
\newline
$ SU(3)|_{supercolor}$.  Because of the selection rules the group
$ SU(3)|_{supercolor}$ is reduced to the subgroup $
SU(2)|$, and therefore the
number of bosons must be 3. Because of the range these bosons must be very
massive. Let us call them ${\aleph}_1$, $\aleph_2$ and $\aleph_3$.
\par We may have an estimation of the strength of the superstrong interaction
in the following way. In ordinary matter(liquids, solids and
gasses) which is formed by the gravitational and electromagnetic forces  the
energy levels are in the eV region. In nuclei which are formed by the
electromagnetic and strong force the energy levels are in the MeV region, and
in quarks which are formed by means of the action of the strong and
superstrong forces the energy leves are in the GeV region. Therefore,
{\it{the superstrong interaction is about a thousand times stronger than the
strong interaction.}}
\par As was said at the beginning of this section the  ideas on prequarks
are very preliminary and a deeper understanding of the
superstrong interaction, as proposed in this work, is under consideration.
This understanding will begin defining the internal quantum numbers of
this interaction.
\vskip .3in
\noindent
VI. THE FUNDAMENTAL INTERACTIONS OF MATTER
\vskip .2in
\hskip 3.5in {\it{A maximis ad minima}}
\vskip .15in
\par It is well known that the electromagnetic interaction is mediated by a
massless vector boson, the photon.  The weak interaction is mediated by the
three heavy vector bosons, $W {+},W {-}$ and $Z{0}$. It was shown by Weinberg,
Salam and Glashow that the weak and electromagnetic interactions are unified at
short distances. The strong force is mediated by the pseudoscalar bosons
$\pi {+}$, $\pi {-}$ and $\pi {0}$(also by $K {0}$, $\bar{K {0}}$ and
$K {\pm}$). In a previous work${(25)}$ it was assumed that the range of
the superweak interaction is infinite. This means that its mediator is a
massless boson.  Let us call this boson the symmetron, $\O$. As has been
shown${(25)}$ the interaction energy of the superweak interaction is of the
form
\begin{eqnarray} V_{12} &=& \frac{\left(A_{I}(2N_{1} - B_{1}) -
A_{B}B_{1}\right)\left(A_{I}(2N_{2} - B_{2}) - A_{B}B_{2}\right)
g {2}}{r}\end{eqnarray}

\noindent
it is positive for like charges. Therefore, the superweak field must be a
vector field. Let us develop the basic equations of the classical superweak
field. Since the boson is massless, we will arrive at the equations by drawing
analogies with electrodynamics. The superweak field is given by

\begin{eqnarray} \varphi &=& \frac{Q}{r}.\end{eqnarray}

\noindent
where $Q = (A_{I}(2N - B) - A_{B}B)g$. The charge $Q$ may be broken into
$Q=Q_{B} + Q_{I}$($I$ of isospin). Because of the variation of $N$ with time,
the charge $Q$ is also a slow function of time governed by the times involved
in the transformation of a nucleon into the other. We can, thus, define a
generalized superweak current density by

\begin{eqnarray} \iota {i} &=& (c\varrho, {\bf{\jmath}})\end{eqnarray}

\noindent
which is actually a baryonic current. Because of baryon number conservation,
for a closed system, what changes with time is $Q_{I}$. Considering that
$v{\ll}c$, ${\bf{\jmath}}$ will be ${\varrho}{\bf{v}}$. In the above formula
$\varrho = \frac{A_{I}(2N - B)g - A_{B}Bg}{V}$, where $V$ is the volume which
contains the charges.  For a fixed $N$ we also find the equation of continuity

\begin{eqnarray} {\bf{\nabla}{\cdot}{\jmath}} +
\frac{\partial{\varrho}}{\partial{t}} &=& 0.\end{eqnarray}

\noindent
We may define a 4-vector field, $\forall {i}$, as

\begin{eqnarray} {\forall}
{i} &=& (\varphi, {\bf{\forall}}).\end{eqnarray}

In this way we may also define the fields ${\bf{\in}}$ and ${\bf{\cap}}$ given
by

\begin{eqnarray} {\bf{\in}} &=& -\frac{1}{c}\frac{\partial\forall}{\partial{t}}
- {\bf{\nabla}}\varphi\end{eqnarray}

\noindent
and
\begin{eqnarray} {\bf{\cap}} &=& {\bf{\nabla}}{\times}{\bf{\forall}}.
\end{eqnarray}

\noindent
With the above fields and using the field strength tensor

\begin{eqnarray} \amalg {\mu\nu} &=& \partial {\mu}{\forall} {\nu} -
\partial {\nu}{\forall} {\mu}\end{eqnarray}

\noindent
we obtain

\begin{eqnarray} \partial_{\mu}(\epsilon {{\mu}\nu{\alpha}\beta}
{{\amalg}_{{\alpha}\beta}}) &=& 0.\end{eqnarray}

\noindent
\par As in electrodynamics we may also define a Lagrangian density for the
field and for the interaction of the field with the superweak charge, given by

\begin{eqnarray} {\cal L} &=& - \frac{1}{16\pi}{\amalg} {\mu\nu}
{\amalg}_{\mu\nu} - \frac{1}{c}\iota_{\alpha}\forall
{\alpha}.\end{eqnarray}

\noindent
In terms of the Lagrangian density we have an action given by

\begin{eqnarray} S &=& \int {\cal L}d {4}x.\end{eqnarray}

\noindent
When we minimize the action $S$, we arrive at

\begin{eqnarray} \partial_{\nu}{\amalg} {\mu\nu} &=&
\frac{4{\pi}}{c}\iota {\mu}.\end{eqnarray}

\noindent
Equations (14) and (17) may be written in terms of the potentials as

\begin{eqnarray} \nabla {2}\varphi +
\frac{1}{c}\frac{\partial}{\partial{t}}({\bf{\nabla}{\cdot}{\forall}}) &=&
- 4\pi\varrho\end{eqnarray}

\noindent
and
\begin{eqnarray} \Box{\bf{\forall} - {\nabla}}\left({\bf{\nabla}
{\cdot}{\forall}} +
\frac{1}{c}\frac{\partial\varphi}{\partial{t}}\right) &=&
- \frac{4\pi}{c}{\bf{\jmath}}.\end{eqnarray}

\noindent
As in electrodynamics the superweak field may be gauged in several ways. If we
choose the Coulomb gauge

\begin{eqnarray} {\bf{\nabla}{\cdot}{\forall}} &=& 0\end{eqnarray}

\noindent
Eq.(18) satisfies the Poisson equation

\begin{eqnarray} \nabla{\varphi} &=& - 4\pi\varrho\end{eqnarray}

\noindent
whose solution is

\begin{eqnarray} \varphi({\bf{x}},t) &=& \int \frac{\varrho({\bf{x}},t)}
{º({\bf{x - x'}})º}d {3}x'.\end{eqnarray}

\noindent
which is an instantaneous potential. Therefore, we may justify the lack of
retarded potential effects in the calculations involving the superweak
potential among galaxies. If we consider that in stars $N$(the number of
neutrons) increases slowly with time and considering the different cycles
of a star we may consider that, for a whole galaxy, $N$ is given by

\begin{eqnarray} N &=& N_{0}(1 - e {-\frac{t}{\tau}}).\end{eqnarray}

\noindent
Since $\tau$ is a very long time, probably, a few bilion years, in
Eq.(22) the approximation

\begin{eqnarray} º{\bf{x - x'}}º &{\ll}& c\tau\end{eqnarray}

\noindent
holds. This means that the static fields may be used.
\par As we showed above the superstrong interaction should be mediated by
three heavy bosons(called mixons). They should have spin 0, and therefore,
the supertrong field should be a scalar field. It will become clear in sections
X, XI and XII that this interaction is responsible for the creation of mass
in quarks and hadrons. Presumably, it may be unified to the gravitational force
at $t=0$. Since we are dealing with a scalar field the Lagrangian density of
the superstrong interaction has the general form

\begin{eqnarray} {\cal L}_{SS} &=&
\frac{1}{2}m{\int} \eta_{\alpha\beta}u {\alpha}u {\beta}\delta {4} x
{\gamma}
- z {\gamma}(\tau)|d\tau \nonumber \\
& & - \frac{1}{8\pi}\left(\phi_{,\alpha}\phi {,\alpha} +
{\mu}\phi\right) -
\mho{\int} d{\tau}\delta {4} x {\alpha} - z
{\alpha}(\tau)|\end{eqnarray}

\noindent
where $z {\alpha}(\tau)$ is the world line of primons. The mass $m$ is
presumably small(or zero) since the superstrong interaction, together with the
strong interaction, are able to generate mass(of quarks and hadrons). If
primons have a small mass, such a mass has to be intrinsic, not caused by the
superstrong interaction. It may be the same kind of mass that leptons have.
The superstrong charge is indicated by $\mho$.
\par The author will not treat the gravitational field in this work. Such a
field seems to be quite elusive. As has been shown$ {(26,27,28)}$ it is
possible to have a scalar field theory of gravity. It is also possible that the
right theory for gravity has not been found yet.
\vskip .3in
\noindent
VII. THE S-MATRIX OF THE SUPERWEAK INTERACTION BETWEEN A NEUTRON AND A PROTON
\vskip .2in
\par Let us, first, calculate the superweak field produced by a proton.
Choosing the Lorentz gauge the superweak potential satisfies Poisson equation

\begin{eqnarray} \Box{\forall} {j}(x) &=& - \frac{4\pi}{c}\iota {j}(x).
\end{eqnarray}

\noindent
In order to find $\iota
{j}$ we need to integrate the above equation. We may
do it following the footsteps of electrodynamics. First, we define a
propagator, $\cal D$ as$ {(29)}$

\begin{eqnarray} {\Box}{\cal D}(x - y) &=& \delta
{4}(x - y)\end{eqnarray}

\noindent
which has the Fourier representation

\begin{eqnarray} {\cal D}(x - y) &=& \int \frac{\delta {4}k}{{2\pi} {4}}
e {-i{\bf{k}{\cdot}(x - y)}}\left(\frac{-1}{k {2} + i\epsilon}
\right)\end{eqnarray}

\noindent
where we have added an infinitesimal positive imaginary part to $k 2$.
\noindent
The solution for the Poisson equation is, therefore,

\begin{eqnarray} {\forall} {j}(x) &=& \int d {4}y{\cal D}(x - y)\iota
{j}(y).
\end{eqnarray}

\noindent
Choosing for $\iota {j}(y)$,

\begin{eqnarray} \iota {j}(y) &=& Q_{p}{\bar{\psi}} {p}_{f}\gamma {j}
{\psi} {p}_{i},\end{eqnarray}

\noindent
where $Q_{p}= - (A_{I} + A_{B})g$, and using plane-wave solutions for free
protons and neutrons, we obtain

\begin{eqnarray} \iota {j}(y) &=& - \frac{(A_{I} + A_{B})g}{V}
\frac{m_{p}}{(E_{i}E_{f})^0.5}e {i(\bf{P}_{f}-\bf{P}_{i}){\cdot}\bf{y}}
\bar{u}(P_{f},S_{f})\gamma {j}u(P_{i},S_{i})\end{eqnarray}

\noindent
and
\begin{eqnarray} S_{fi} &=& \frac{i4\pi({A_{I}} {2} - {A_{B}} {2})g
{2}}{cV 2}
(2\pi) {4}\delta {4}(P_{f} - P_{i} + p_{f} - p_{i})\frac{m_{n}}
{\sqrt{E_{f} {n}E_{i} {n}}}\frac{m_{p}}{\sqrt{E_{f} {p}E_{i}
{p}}} \nonumber \\
& & {\times}\left(\bar{u}(p_{f},s_{f})\gamma_{j}u(p_{i},s_{i})\right)
\frac{1}{(p_{f} - p_{i})
{2} + i\epsilon}\left(\bar{u}(P_{f},S_{f})\gamma {j}
u(P_{i},S_{i})\right).\end{eqnarray}

\noindent
The above S-matrix element corresponds to the Feynman graph shown below.
Therefore, the cross section of the superweak interaction between two protons
is $(A_{I} - A_{B})g {2}/e
{2}$ smaller than the electromagnetic interaction
between them.

\vskip 3.5in
\noindent
VIII. THE UNIFICATION OF THE SUPERWEAK AND STRONG INTERACTIONS
\vskip .2in
\par As was shown in section VI the superweak field satisfies the Poisson
equation

\begin{eqnarray} \Box{\forall {\alpha}} &=& - \frac{4\pi}{c}\iota
{\alpha}.
\end{eqnarray}

\noindent
As has been shown$ {(25)}$, as $t{\rightarrow}0$, $v{\rightarrow}0$ and
$N{\rightarrow}Z$(or $\eta{\rightarrow}0.5$). Therefore,
$\iota {\alpha}{\rightarrow}\iota {o}= - A_{B}Bg/V$ and
$\forall {\alpha}{\rightarrow}\forall
{o}= \varphi$. Thus, Eq.(33) becomes

\begin{eqnarray} \Box{\varphi} &=& 4{\pi}\frac{A_{B}Bg}{cV}.\end{eqnarray}

\noindent
Making the substitution $\varphi'=\frac{cV}{A_{B}B}\varphi$, one obtains

\begin{eqnarray} \Box{\varphi'} &=& 4{\pi}g\end{eqnarray}

\noindent
whose solution, for small $r$, is indistinguishable from the solution of
the equation

\begin{eqnarray} \Box{\varphi'} - {\mu}\varphi' &=& 4{\pi}g.\end{eqnarray}

\noindent
{\it{Therefore, the superweak interaction is unified to the strong interaction
at $t=0$.}}\par Therefore, resuming the results we are able to construct the
fol
table for the interactions of matter:
\vskip .3in
\centerline{TABLE OF THE UNIFICATION OF THE FORCES OF NATURE}
\vskip .25in
\centerline{Weak(3 heavy vector bosons) with Electromagnetic(1 massless vector
boson)}
\centerline{Strong(3 heavy pseudoscalar bosons) with Superweak(1 vector
massless
boson)}
\centerline{Superstrong(3 heavy scalar bosons) with Gravity(1 massless
boson) ???}
\vskip .25in
\par From the above table we may imply that the unitary group of gravity has to
be $U(1)$ and that the corresponding boson is massless. A summary of the
fundamental interactions of nature is presented in Table 6.
\vskip .3in
\noindent
IX. QUARK CONFINEMENT
\vskip .2in
\par Quantum chromodynamics(QCD) has been quite successful at providing a
simple picture of many processes involving hadrons and leptons. However, many
features of QCD remain unanswered. Two of them are quark confinement and the
energies of hadrons, including the resonant states. Also, the number of quarks
is so high that we may ask if they are elementary, after all. Moreover, QCD
lacks a dynamical content.
\par It has been proposed above that quarks are composed of prequarks which
are true elementary particles, together with leptons. It has been shown above
that two forces act in the interaction among quarks. These two forces
are the strong and the superstrong forces. It has also been suggested
that these forces must have opposite signs when acting among quarks.
\par Present experiments show that quarks are permanently confined inside
hadrons, at least, within the energies allowed by the present generation of
particle accelerators.
\par Quark confinement can be explained, based on first principles,  as the
result of the superstrong
and strong interactions together. Because the quark is a structured state it
is formed by the action of the superstrong and strong forces. That is, we
expect that there should exist an effective attractive potential  as we have in
the other kinds of structured matter. Therefore, we also  expect to have a
sort of Lennard-Jones effective potential in the interaction among quarks.
Expanding a Lennard-Jones potential around the minimum we obtain a harmonic
oscillator potential.  Thus, if we consider that in their lowest state of
energy quarks are separated by a distance $r_{q}$, then for small departures
from equilibrium the potential must be of the form
\begin{eqnarray} V(r) &=& V_{o} + \frac{K}{2}(r - r_{q})
{2}\end{eqnarray}\

\noindent
where $K$ is a constant and $V_{o}$ is a negative constant representing the
depth of the potential well. As $r$ increases the restoring force among
quarks also increases. Because of it quarks may be permanently confined.
\vskip .3in
\noindent
X. THE ENERGIES OF BARYON STATES(INCLUDING ALL KNOWN RESONANCES)
\vskip .2in
\par Around its minimum we may approximate a Lennard-Jones potential by the
potential of a harmonic oscillator and include the anharmonicity as a
perturbation. By doing so we may be able to calculate the energies of almost
all baryon states.
\par Let us consider a system composed of three quarks which interact in pairs
by means of a harmonic potential. Let us disregard the electromagnetic
interaction which must be considered as a perturbation. Also, let us  disregard
any rotational contribution which may enter as a perturbation too. This is
reasonable because the strong and superwtrong interactions must be much larger
than the ``centrifugal'' potential. If we consider that quarks do not move
at relativistic speeds, and disregarding  the spin interaction
among quarks, we may just use Schr\"{o}dinger equation in terms of normal
coordinates$ {(30)}$

\begin{eqnarray} \sum_{i=1} {6}
\frac{{\partial} {2}\psi}{{\partial}{\xi}_{i}
{2}} +
\frac{2}{{\hbar} 2}\left(E -
\frac{1}{2}\sum_{i=1} {6}{\omega}_{i}{{\xi_i}
2}\right)\psi &=& 0\end{eqnarray}

\noindent
where we have used the fact that the three quarks are always in a plane. The
above equation may be resolved into a sum of 6 equations

\begin{eqnarray} \frac{{\partial} {2}\psi}{{\partial}{\xi}_{i} {2}} +
\frac{2}{{\hbar} 2}\left(E_{i} - \frac{1}{2}\omega_{i}{\xi_i}
{2}\right)\psi
&=& 0,\end{eqnarray}

\noindent
which is the equation of a single harmonic oscillator of potential energy
$\frac{1}{2}\omega_{i}{\xi_i} {2}$ and unitary mass with

\begin{eqnarray} E &=& \sum_{i=1}
{6} E_{i}.\end{eqnarray}

\noindent
The general solution is a superposition of 6 harmonic motions in the 6 normal
coordinates.
\par The eigenfunctions $\psi_{i}(\xi_i)$ are the ordinary harmonic oscillator
eigenfuntions

\begin{eqnarray} \psi_{i}(\xi_{i}) &=& N_{v_i}e {-(\alpha_{i}/2)\xi_{i}
{2}}
H_{v_i}(\sqrt{\alpha_{i}}\xi_{i}),\end{eqnarray}

\noindent
where $N_{v_i}$ is a normalization constant, $\alpha_{i} = \nu_{i}/{\hbar}$ and
$H_{v_i}(\sqrt{\alpha_{i}}\xi_{i})$ is a Hermite polynomial of the $v_i$th
degree. For large $\xi_{i}$ the eigenfunctions are governed by the exponential
functions which make the eigenfunctions go to zero very fast.  Of course,
{\it{this is valid for any energy and must be the reason behind quark
confinement}}. We will come back yet to this point after calculating the
possible energy levels of baryons.
\par The energy of each harmonic oscillator is

\begin{eqnarray} E_{i} &=& h\nu_{i}(v_{i} + \frac{1}{2}),\end{eqnarray}

\noindent
where $v_{i} = 0,1,2,3,...$ and $\nu_i$ is the classical oscillation frequency
of the normal ``vibration'' $i$, and $v_i$ is the ``vibrational'' quantum
number. The total energy of the system can assume only the values

\begin{eqnarray} E(v_{1},v_{2},v_{3}, ...v_{6}) &=& h\nu_{1}(v_{1} +
\frac{1}{2}) + h\nu_{2}(v_{2} + \frac{1}{2}) + ... h\nu_{6}(v_{6} +
\frac{1}{2}).\end{eqnarray}

\par As was said above the three quarks in a baryon must always be in a plane.
Therefore, each quark is composed of two oscillators and so we may rearrange
the energy expression as

\begin{eqnarray} E(n,m,k) &=& h\nu_{1}(n + 1) + h\nu_{2}(m + 1) +
h\nu_{3}(k + 1),\end{eqnarray}

\noindent
where $n=v_{1} + v_{2},m=v_{3} + v_{4},k=v_{5} + v_{6}$. Of course, $n,m,k$ can
assume the values, 0,1,2,3,... We may find the constants $h\nu$ from the ground
states of some baryons. They are the known quark masses taken as
$m_{u}=m_{d}= 0.31$Gev, $m_{s}= 0.5$Gev, $m_{c}=1.7$Gev and $m_{b}=5$Gev. The
mass of the top quark has not been determined yet, but as we will show the
present theory may help in the search for its mass.

\par Let us start the calculation with the states ddu(neutron), uud(proton) and
ddd($\Delta {-}$), uuu ($\Delta
{++}$) and their resonances. All the energies
below are given in Gev. The experimental values of baryon masses were taken
from reference 31. Because $m_{u}=m_{d}$, we have that the energies calculated
by the formula

\begin{eqnarray} E_{n,m,k} &=& 0.31(n+m+k + 3)\end{eqnarray}

\noindent
correspond to many energy states. Some levels contain an ``intrinsic'' pion
excitation and their energies are the sum of the energy of the previous level
plus a pion energy. This happens whenever the difference between any two levels
is of the order of 280MeV. This fact induces, somehow, one or more levels
between the two levels. Of course, this is completely ad hoc at this point and
needs further investigation. Some levels may also be caused by the three-fold
degeneracy which may be raised if we take into account the anharmonicity of the
potential. As we can conclude from the experimental values this contributes
roughly a pion mass. Further splitting is expected(as with the other regular
levels $E_{n,m,k}$) due to electromagnetic and weak interactions. The state
$E_{0,0,0} + \pi$ should not be allowed, somehow, because there is no level
located at 1.07Gev. The justification for the omission of this level will have
to wait for the understanding of the superstrong interaction, but tt may be
related to the stability of the proton which is the state $E_{0,0,0}$. The
calculated values are displayed in Table 7.
\par For the $\Xi$ particle the energies are expressed by

\begin{eqnarray} E_{n,m,k} &=& 0.31(n+1) + 0.5(m+k+2).\end{eqnarray}

\noindent
See Table 8 to check the agreement with the experimental data.
\par In the same way the energies of $\Omega$ are obtained by

\begin{eqnarray} E_{n,m,k} &=& 0.5(n+m+k+3).\end{eqnarray}

\noindent
Again, there are energies given by $E_{n,m,k} + \pi$. The energies are
displayed in Table 9. The discrepancies are higher, of the order of 10\%
and decreases as the energy increases. This is a tendency which is also
observed for the other particles. This may mean that, at the bottom, the
potential is flatter than the potential of harmonic oscillator. That is, the
effective potential must have terms of fourth order or higher order terms.
As the energy increases the potential becomes closer to the one of a harmonic
oscillator.
\par In the same fashion we list below the formulas for calculating the
energies of many other states. The energies of the charmed baryons($C=+1$)
$\Lambda_{c} {+}$, $\Sigma_{c} {++}$, $\Sigma_{c} {+}$ and $\Sigma_{c}
{0}$
are given by

\begin{eqnarray} E_{n,m,k} &=& 0.31(n+m+2) + 1.7(k+1).\end{eqnarray}

\noindent
The levels are shown on Table 10.
For the charmed baryons($C=+1$) $\Xi_{c} {+}$ and $\Xi_{c} {0}$ we have

\begin{eqnarray} E_{n,m,k} &=& 0.31(n+1) + 0.5(m+1) + 1.7(k+1).\end{eqnarray}

\noindent
The results are displayed on Table 11. As for the $\Omega_{c} {0}$, its
energies are

\begin{eqnarray} E_{n,m,k} &=& 0.5(n+m+2) + 1.7(k+1).\end{eqnarray}

\noindent
Table 12 shows the energy levels results.
\par We may predict the energies of many particles given by the formulas:
\begin{itemize}
\item ucc and dcc, $E_{n,m,k} = 0.31(n+1) + 1.7(m+k+2)$;
\item scc, $E_{n,m,k} = 0.5(n+1) + 1.7(m+k+2)$;
\item ccc, $E_{n,m,k} = 1.7(n+m+k+3)$;
\item ccb, $E_{n,m,k} = 1.7(n+m+2) + 5(k+1)$;
\item cbb, $E_{n,m,k} = 1.7(n+1) + 5(m+k+2)$;
\item ubb and dbb, $E_{n,m,k} = 0.31(n+1) + 5(m+k+2)$ ;
\item uub and ddb, $E_{n,m,k} = 0.31(n+m+2) + 5(k+1)$;
\item bbb, $E_{n,m,k} = 5(n+m+k+3)$.
\end{itemize}

\par We clearly see from Tables 7,8,9,10,11,12 and 13 that there are minor
discrepancies in some states between the calculated energy and the
experimental data. This is expected because of the assumptions that we made.
An improved model must include the electromagnetic interaction, the effect
of rotation(or spin) and the isospin. All of them will split the levels.
Also, we must include the anharmonicity of the potential.
The present treatment assumed that the oscillators are independent. But, this
may not be the case and there must exist coupling between them.
\par From the above discussion we see clearly that it is meaningless to
try to make quarks free by using more energetic collisions  between
two baryons because we will just get more excited states. That is, we will
obtain more and more resonances. This is so because, as as shown above,
as the separation between two quarks increases the wavefunction vanishes very
fast regardless of the state of energy.
\par According to our considerations the total mass of a baryon must be given
by
\begin{eqnarray} M &=& M_{K} + V_{S,SS} + V_{e} + V_{spin} \end{eqnarray}

\noindent
where $M_{K}$ is the total kinetic energy of the three constituent quarks,
$V_{S,SS}$ is the potential of the combination of the strong and superstrong
interactions(our Lennard-Jones potential), $V_{e}$ is the electromagnetic
interaction, and $V_{spin}$ is the spin dependent term of the mass.  We showed
above that $V_{S,SS}$ is the leading term. Since quarks have charges the term
$V_{e}$ must contribute in the splitting of the levels. Therefore, we conclude
that $M_{K}$ must be very small. As is argued by Lichtenberg${32}$ it is hard
to see how $SU(6)$ is a good approximate dynamical symmetry of baryons if
quarks move at relativistic velocities inside baryons. As has been shown by
Morpurgo${33}$ a nonrelativistic approximation might be a good one. Our
treatment above agrees well with this conclusion since we used
Schr\"{o}dinger's equation for the oscillators.
\par We see that the masses of baryons are described quite well by
the simple model above described. It lends support to the general framework of
having quarks as the basic building particles of baryons. Therefore, it agrees
well with QCD. But, the model is also based on the idea of having a
substructure for quarks, i.e., a superstrong interaction which may be the
interaction responsible for the masses of quarks. We also have shown that
quarks may be permanently confined inside baryons. The model needs, of course,
further developments in order to include the electromagnetic interaction and
the contributions of the spin and isospin to the mass.
\vskip .3in
\noindent
XI. THE EXCITED STATES OF QUARKS
\vskip .2in
\par The interaction between two primons brings forth the action of the strong
force between them, and there is also the action of the superstrong interaction
between them. Therefore, there is a sort of effective potential energy which
can be approximated by a Lennard-Jones potential energy. This potential energy
is harmonic around its minimum.  Considering that primons do not move at
relativistic speeds and doing in the same fashion as we did in section X we
obtain that the masses of quarks should be given by

\begin{eqnarray} E_{q} &=& \bar{\hbar}\nu(k + 1/2)\end{eqnarray}

\noindent
in which $k = 0,1,2,3,...$ and $\bar{\hbar}$ has been previously defined.
\par Let us now determine the energy levels of the quarks u,d,s,c,t and b. The
ground state of the u quark is about 0.31GeV. Therefore,
$\bar{\hbar}\nu=0.62$GeV. Thus, the energies of the $u$ and $d$ quarks are
0.31GeV, 0.93GeV, 1.55GeV, 2.17GeV, ... Making the same for the other quarks we
obtain the following results:
\begin{itemize}
\item $E_{u} = E_{d} = 0.62(k + 1/2)$;
\item $E_{s} = 1(k + 1/2)$;
\item $E_{c} = 3.4(k + 1/2)$;
\item $E_{b} = 10(k + 1/2)$;
\item $E_{t} = \bar{\hbar}\nu_{t}(k + 1/2)$.
\end{itemize}

\par Since $k + 1/2$ is a half integer, we may also write $E_{q}$ as

\begin{eqnarray} E_{q} &=& E_{o}l\end{eqnarray}

\noindent
where $l$ is an odd integer. The calculated values are shown on Table 14.
\par Due to the overtones of the vibrations in the effective potential energy
there must also exist other energy levels. For example, the overtones of the
ground state with all the other excited states for the $u$ and
$d$ quarks produce the energy levels 1.24GeV, 1.86GeV, 2.48GeV, etc.  In the
same way for the $s$ quark one obtains 2.0GeV, 3.0GeV, 4.0GeV, etc. We can
easily see that the higher overtones coincide either with the regular levels
or with the overtones of the ground state.
\par Because quarks are confined we could observe the
excited levels only by means of hadrons, but actually we can not. The reason
is simple. Let us, for example, consider the baryons made of the first
excited states of the u or d quarks. According to Eq.(44) such baryons have
energies

\begin{eqnarray} E {*}_{111u} &=& E
{*}_{111d} = 0.93(n + 1) + 0.93(m + 1) +
0.93(k + 1) \nonumber \\
&=& 0.93(n + m + k + 3)\end{eqnarray}

\noindent
in which the asterisk means excited state and the repeated ones mean that each
quark is in the first excited state.  But $0.93(n + m + k + 3) = 0.31(3(n +
m + k) + 9)$ can be represented as $0.31(n' + m' + k' + 3)$, where $n', m'$,
and $k'$ are also integers that satisfy the relation

\begin{eqnarray} n' + m' + k' &=& 3(n + m + k + 2).\end{eqnarray}

\noindent
\par Therefore, the baryons made of quark excited states cannot be observed
because their energy levels coincide with the energies of baryons made of
quark ground states. The same also happens for the overtones.
\vskip .3in
\noindent
XII. THE ENERGIES OF MESONS
\vskip .2in
\par According to QCD a meson is a colorless state which transforms  under
$SU_{3}$ as

\begin{eqnarray} q {in}q_{jn} &=& \bar{q}_{in}q_{jn}.\end{eqnarray}

\noindent
According to the theory presented above it is reasonable to admit that there is
also a harmonic effective potential energy in the interaction between a quark
and an antiquark.  As we know they are also confined inside mesons, which,
actually, supports the ideas above. In this fashion the energies of many mesons
should be given by

\begin{eqnarray} E_{n} &=& \hbar\nu(n + 1/2).\end{eqnarray}

\noindent
Let us apply it to the pion family. The ground state must
correspond to the three pions $\pi {+}$, $\pi {-}$ and $\pi
{0}$. The splitting
comes from the electromagnetic interaction(and from the anharmonicity of the
potential energy). What to choose for ${\hbar}\nu_{\pi}$? We know that it is in
the range 270-280 MeV. The values of the corresponding
energies are shown in Table 15. The levels have been labeled as $\pi_{n}$.
There is a good agreement for almost all levels(error below 3\%). The level
with energy between 405 and 420 MeV is forbidden, somehow.
The experimental values were taken from reference 31, except the energy of the
$\epsilon$ meson, which was taken from reference 34. We may predict that there
must exist mesons of the pion family with energies between 2565 and 2660MeV.
Of course, many levels should come from the inumerous overtones of the ground
state with all the other excited states. For example, the $f_{o}$(1400MeV)
meson is just the overtone $\pi_{0+4}$. All the other levels can be identified
in the same way. The results are summarized on Table 16.
\par The energies of the other mesons are calculated in the following way:
There are levels, called regular levels, which follow Eq. (57), and there are
levels that are overtones of the regular levels with the pion levels. This
means that the net potential(of the superstrong and strong interaction) also
generates pion mass oscillations. This is justified in quantum mechanics if
we assume that the hamiltonian $H$ can be broken in two parts, one of the
given meson and the other of the pion, i.e.,

\begin{eqnarray} H\Psi &=& (H_{M} + H_{\pi})\Psi = E\Psi = (E_{M} +
E_{\pi})\Psi\end{eqnarray}

\noindent
where $H_{M}$ is the hamiltonian of a given meson and $H_{\pi}$ is the
hamiltonian of the pion. The wavefunction $\Psi = \Psi_{M}\Psi_{\pi}$, in
which $\Psi_{M}$ is the meson wavefunction and $\Psi_{\pi}$ is the wavefuntion
of a pion state. Therefore, the energy $E$ is just

\begin{eqnarray} E &=& \hbar\nu_{M}(n + \frac{1}{2}) + \hbar\nu_{\pi}(k +
\frac{1}{2}).\end{eqnarray}

\noindent
Adding the overtones of the pion family we also have that many levels will be
given by

\begin{eqnarray} E &=& \hbar\nu_{M}(n + \frac{1}{2}) + \hbar\nu_{\pi}(k +
\frac{1}{2}) + \hbar\nu_{\pi}(m + \frac{1}{2}).\end{eqnarray}

\noindent
The first levels are for $n=0$, $m=0$, and any value of $k$.
We should keep in mind that any wavefunction that we are considering is the
spatial part of the total wavefunction. The other part of it is due to
isospin and has not been considered in this work. For the $K$ meson let us take
$\hbar\nu_{K}$=996MeV. The regular energy levels, $K_{n}$ are 498MeV, 1494MeV,
2490MeV, etc. Many other states are the overtones of pion states with $K_{0}$.
These results mean that the excited states of kaons are not pure $\bar{s}u$,
$\bar{s}d$ states, but should also contain pion states, which here are
represented by $\bar{u}u$. The results are listed on Tables 17 and 18.
For energies higher than about 1629MeV we have the overtones of $K_{1}$ with
the pion states. These have not been included in the tables because it becomes
very difficult to know if a given state belongs to first array of overtones or
to the second one. This will have to wait for the knowledge of the selection
rules that we should follow.
\par Let us, now, consider the $D$ mesons. We have $\hbar\nu_{D} = 3728$MeV.
The regular levels according to Eq.(134) are 1864MeV, 5592MeV,
9320MeV, etc.The ground state overtones are 7456MeV, 11184MeV, etc. The
observed states $D {\ast}$, $D_{1}$ and $D{\ast}_{2}$ are just overtones of
the ground states with pion states. That is, $D{\ast}(2010) = D_{0}(1864)
+ \pi_{0}(140)$; $D_{1}(2420) = D_{0}(1864) + \pi_{0+1}(540-560)$.
Tables 19 and 20 show these results.
\par In the same fashion for $D_{S}$ mesons, $\hbar\nu_{D_{S}} = 3938$MeV.
The regular energy levels are 1968MeV, 5907MeV, 9845MeV, etc. The ground state
overtones are 7875MeV, 11813MeV, etc. The states $D {\ast}_{S}$ and
$D_{S1}$ are overtones of the ground state with pion states, i.e.,
$D {\ast}_{S}(2110) = D_{S(0)}(1969) + \pi_{0}(140)$;
$D_{S1}(2536) = D_{S(0)}(1969) + \pi_{0+1}(540-560)$. The levels are found
on Tables 21 and 22.
\par Only the ground state of the $B$ meson has been observed experimentally.
Expecting that it follows the same general trend shown above we may predict
the possible excited states(Tables 23 and 24).
\par The $c\bar{c}$ mesons may be treated in the same way if we consider that
$\eta_{c}$(2980) is the ground state. Table 25 shows the regular states
calculated according to $\hbar\nu_{c\bar{c}}(n + 0.5)$ with
$\hbar\nu_{c\bar{c}}=5960$MeV. The other excited states are the overtones of
$\eta_{c}(2980)$ with pion states. For instance, $\chi_{c0}(3415)$ is the
overtone $\eta_{c}(2980) + \pi_{1}(405-420)$. In the same fashion we have that
$\chi_{c2}(3556) = \eta_{c}(2980) + \pi_{0+1}(540-560)$. The calculated values
are listed on Table 26.
\par The energies of the regular states of $b\bar{b}$ are given by
$\hbar\nu_{b\bar{b}}(n + 0.5)$ with $\hbar\nu_{b\bar{b}}=18920$. The other
excited states of the $b\bar{b}$ mesons follow the same trend of the heavy
mesons, that is, they are just overtones of the ground state
$\Upsilon(9460)$ with pion states. One state(and possibly others to be found),
however, is a superposition of the ground state with an excited state of the
kaon. It is the state $\Upsilon(10355) = \Upsilon(9460) + K_{0}(498) +
\pi_{1}(405-420)$. Some examples of the other states are
$\chi_{b0}(9860) = \Upsilon(9460) + \pi_{1}(405-420)$;
$\Upsilon(10023) = \Upsilon(9460) + \pi_{0+1}(540-560)$;
$\chi_{b2}(10270) = \Upsilon(9460) + \pi_{0+2}(810-840)$. All states are listed
on Tables 27 and 28.
\par Of course, for all mesons we may predict the possible values of the
energies of many levels to be found experimentally. We observe that some
overtones do not occur. At this point there is no explanation for this fact. Of
course, it may be so because of selection rules yet to be found.
\par According to the above theory the largest contribution to the masses of
hadrons(and of quarks) comes from the strong and superstrong
interactions(together). But, what about the energies of leptons?
We can immediately see that the masses of the electon, muon and $\tau$ do not
correspond to the levels of a harmonic oscillator.   This is in a certain way
in agreement with what was said in the beginning of this work, because we
found that leptons should belong to a separate class and must be
elementay particles, just as primons are. If this theory is true the Higgs
scalar field does not exist at all. All experiments up to now show exactly
this.
Its mass has been searched in many different ranges and has not been found.
As this paper shows we will have to modify QCD, by including the dynamical part
that it lacks.
\par In the light of what was shown above, we can understand the different
decay channels of mesons into kaons or into pions. For example, the meson
$a_{0}(983)$ is either $K_{0} + \pi_{0 + 1}$ or $\pi_{3}(945-980)$. Therefore,
we expect that it will decay into kaons or into pions. In the same way, we
expect $f_{0}$ to decay as $K\bar{K}$ and as $\eta\rho$. The $f_{2}(1270)$
meson may be either $K_{0} + \pi_{0 + 2}$ or $\pi_{4}(1215-1260)$. Thus, it
will decay into $\pi\pi$ and $K\bar{K}$. Its decay $\eta\eta$ is a consequence
of the decay $\pi\pi$. This is also the case for many other mesons. According
to Table 18 $\omega(1594)$(or another resonance with about the same energy)
will also have a decay mode involving kaons.
\vskip .3in
\noindent
XIII. CONCLUSION
\vskip .2in
\par It is proposed that nature has six fundamental forces which are unified in
pairs and, therefore, reduced to three at $t=0$. Some general ideas concerning
the characteristics of the superstrong interaction have been presented.
The superweak interaction is responsible for the expansion and contraction of
the Universe. It is shown that it is unified with the strong force at $t=0$.
\par A reasonable physical explanation for quark confinement is provided.
The energies of all hadrons including all known resonances are calculated in
a simple manner. The energies of all other hadrons to be discovered are given.
The excited states of quarks have also been calculated and has been shown why
they can not be observed. The paper predicts the existence of three heavy
bosons of the superstrong interaction and a massless boson of the superweak
interaction. It is proposed that the gauge group of gravity is $U(1)$.
\par The evidence of the superstrong interaction is obvious
from the calculated values of the energies of hadrons. The non-existence of
the Higgs boson also gives support to the existence of a superstrong
interaction.
\par It has been shown that the superweak force is unified to the strong force
at $t=0$.
\vskip .3in
\noindent
\Large{References}
\normalsize
\vskip .3in
\noindent
1. A. A. Penzias and R. W. Wilson, {\it{Astrophys. J.}} {\bf{142}}, 419(1965).
\newline
2. J. C. Mather, Cheng,E.S., Eplee,R.E., Isaacman,R.B., Meyer,S.S.,
Shafer,R.A., Weiss,R., Wright,E.L., Bennet,C.L., Boggess,N.W., Dwek,E.,
Gulkis,S., Hauser,M.G., Janssen,M., Kelsall,T., Lubin,P.M., Moseley,Jr.,S.H.,
 Murdock,T.L., Siverberger,R.F., Smoot,G.F.,and Wilkinson,D.T., {\it{Astrophys.
J.(Letters)}} {\bf{354}}, L37(1990).
\newline
3. E. Hubble, in {\it{Proceedings of the National Academy of Science}},
{\bf{15}},168-173(1929).
\newline
4. J. C. Mather, talk at the XIII Interantional Conference on
General Relativity and Gravitation, Huerta Grande, Argentina, 1992.
\newline
5. V. de Lapparent, M. J. Geller, and J. P. Huchra, {\it{Astrophys.
J.(Letters)}} {\bf{302}}, L1(1986).
\newline
6. T. J. Broadhurst, R. S. Ellis, D. C. Koo, and A. S. Szalay, {\it{Nature}}
{\bf{343}},726(1990).
\newline
7. A. Dressler and S. M. Faber, {\it{Astrophys. J.}} {\bf{354}}, 13(1990).
\newline
8. A. Dressler, S. M. Faber, D. Burnstein, R. L. Davies, D. Lynden-Bell,
R. J. Terlevich, and G. Wegner, {\it{Astrophys. J.(Letters)}} {\bf{313}},
L37(19
\newline
9. D. Lynden-Bell, S. M. Faber, D. Burnstein, R. L. Davies, A. Dressler,
R. J. Terlevich, and G. Wegner, {\it{Astrophys. J.}} {\bf{326}}, 19(1988).
\newline
10. H. Kurki-Suonio, G. J. Mathews, and G. M. Fuller, {\it{Astrophys.
J.(Letters)}} {\bf{356}}, L5(1990).
\newline
11. S. J. Maddox, G. Efstathiou, W. J. Sutherland and J. Loveday, {\it{Mon.
Not.Astr.Soc.}} {\bf{242}}, 43p(1990).
\newline
12. W. Saunders. M. Rowan-Robinson, and A. Lawrence, ``The Spatial Correlation
Function of IRAS Galaxies on Small and Intermediate Scales'', 1992, QMW
preprint.
\newline
13. G. B. Dalton, G. Efstathiou, S. J. Maddox, and W. J. Sutherland,
{\it{Astrophys. J. Lett.}} {\bf{390}}, L1, 1992.
\newline
14. M. S. Vogeley, C. Park, M. J. Geller, and J. P. Huchra, {\it{Astrophys.
J. Lett.}} {\bf{391}}, L5, 1992.
\newline
15. C. Park, J. R. Gott, and L. N. da Costa, {\it{Astrophys.  J. Lett.}}
{\bf{392}}, L51, 1992.
\newline
16. M. Rowan-Robinson, {\it{New Scientist}} {\bf{1759}}, 30(1991).
\newline
17. W. Saunders, C. Frenk, M. Rowan-Robinson, G. Efstathiou, A. Lawrence,
N. Kaiser, R. Ellis, J. Crawford, X.-Yang Xia and I. Parry, {\it{Nature}}
{\bf{349}}, 32(1991).
\newline
18. S. White, talk at the XIII International Conference on General
Relativity and Gravitation, Huerta Grande, Argentina, 1992.
\newline
19. H. Fritzsch, in {\it{Proceedings of the twenty-second Course of the
International School of Subnuclear Physics, 1984}}, ed. by A. Zichichi
(Plenum Press, New York, 1988).
\newline
20. W. K\"{u}pper, G. Wegmann, and E. R. Hilf, {\it{Ann. Phys.}} {\bf{88}},
454(
\newline
21. D. Bandyopadhyay, J. N. De, S. K. Samaddar, and D. Sperber, {\it{Phys.
Lett. B}} {\bf{218}}, 391.
\newline
22. J. E. Lennard-Jones, {\it{Proc. Roy. Soc.}} {\bf{A106}}, 463(1924).
\newline
23. K. Huang, in {\it{Quarks, Leptons and Gauge Fields}}(World Scientific,
Singapore, 1982).
\newline
24. G. 'tHooft, in {\it{Recent Developments in Gauge Theories}}, eds. G.
'tHooft et al.(Plenum Press, New York, 1980).
\newline
25. M. E. de Souza, in {\it{On the General Properties of Matter}}, report
no. MES-02-092493, September 1993.
\newline
26. C. Brans and R. H. Dicke, {\it{Phys. Rev.}} {\bf{124}}, 925(1961).
\newline
27. P. Jordan, {\it{Z. Phys.}} {\bf{157}}, 112(1959).
\newline
28. W.-T. Ni, {\it{Astrophys. J.}} {\bf{176}}, 769(1972).
\newline
29. J. D. Bjorken and S. D. Drell, in {\it{Relativistic Quantum Mechanics}}
(McGraw-Hill, New York, 1964).
\newline
30. L. Pauling and E. B. Wilson Jr., {\it{Introduction to Quantum Mechanics}}
(McGraw-Hill, New York, 1935).
\newline
31. Particle Data Group, {\it{Review of Particle Properties}}, Phys. Rev. D,
{\bf{45}}, Part II, No. 11 (1992).
\newline
32. Lichtenberg, D. B., {\it{Unitary Symmetry and Elementary Particles}}
(Academic Press, New York, N.Y) 1970.
\newline
33. Morpurgo, G., {\it{14th International Conference on High Energy Physics}},
Vienna, p.225. CERN, Geneva.
\newline
34. {\it{Review of Particles Properties}}, Rev. Modern Phys., {\bf{45}},
Supplem., 1973.
\newpage

\begin{center}
\begin{tabular}{c c c c c c c} \hline\hline\\
& ? & & quark & & nucleon & \\
\\
& nucleon & & nucleus & & atom & \\
\\
& atom & & gas & & galaxy & \\
& & & liquid & & & \\
& & & solid & & & \\
\\
& galaxy & & galactic liquid & & ? \\
\\
\hline\hline\\
\end{tabular}
\end{center}
\vskip .5in

\begin{center}
\parbox{4in}
{Table 1. The two general states  which make everything in the
Universe, stepwise. The table is arranged in such a way to show the links
between the polarized states and the single states.}
\end{center}

\newpage

\begin{center}
\begin{tabular}{c c c} \hline\hline\\ ? & ? & strong force \\
& strong force & \\
\\
strong force & strong force & electromagnetic force \\
& electromagnetic force & \\
\\
electromagnetic force & electromagnetic force & gravitational force \\
& gravitational force & \\
\\
gravitational force & gravitational force & ?\\
& ? & \\
\\
\hline\hline\\
\end{tabular}
\end{center}
\vskip .5in

\begin{center}
\parbox{5in}
{Table 2. Three of the fundamental forces of nature. Each force
appears twice and is linked to another force by means of a correlated
state. The interrogation marks suggest that there should exist two other
forces. Compare with Table 1.}
\end{center}
\newpage

\begin{center}
\begin{tabular}{  cc  ccccc } \hline\hline
& & & & & & \\
superflavor & & & charge & & spin & \\
& & & & & & \\
\hline \hline
& & & & & & \\
$p_{1}$ & & & $\frac{5}{6}$  & & $\frac{1}{2}$ & \\
& & & & & & \\
\hline
& & & & & & \\
$p_{2}$ & & & $-\frac{1}{6}$ & & $\frac{1}{2}$ & \\
& & & & & & \\
\hline
& & & & & & \\
$p_{3}$ & & & $-\frac{1}{6}$ & & $\frac{1}{2}$ & \\
& & & & & & \\
\hline
& & & & & & \\
$p_{4}$ & & & $-\frac{1}{6}$ & & $\frac{1}{2}$ & \\
& & & & & & \\
\hline\hline
\end{tabular}
\end{center}
\vskip .5in

\begin{center}
\parbox{4in}
{Table 3. Table of charges and spins of primons.}
\end{center}

\newpage

\begin{center}
\begin{tabular}{  cc  ccc ccc ccc ccc  } \hline\hline
& & & & & & & & & & & & &  \\
 & & & $p_{1}$ & & & $p_{2}$ & & & $p_{3}$ & & & $p_{4}$ &  \\
& & & & & & & & & & & & &  \\
\hline\hline
& & & & & & & & & & & & &  \\
$p_{1}$ & & &   & & & u & & & s & & & t &  \\
& & & & & & & & & & & & &  \\
\hline
& & & & & & & & & & & & &  \\
$p_{2}$ & & & u & & &   & & & d & & & c &  \\
& & & & & & & & & & & & &  \\
\hline
& & & & & & & & & & & & &  \\
$p_{3}$ & & & s & & & d & & &   & & & b &  \\
& & & & & & & & & & & & &  \\
\hline
& & & & & & & & & & & & &  \\
$p_{4}$ & & & t & & & c & & & b & & &   &  \\
& & & & & & & & & & & & &  \\
\hline\hline
\end{tabular}
\end{center}
\vskip .5in

\begin{center}
\parbox{4in}
{Table 4. Table of composition of quark flavors.}
\end{center}

\newpage

\begin{center}
\begin{tabular}{  ccc  ccc ccc ccc  } \hline\hline
& & & & & & & & & & & \\
&          & & &  $\alpha$ & & & $\beta$ & & & $\gamma$ & \\
& & & & & & & & & & & \\
\hline\hline
& & & & & & & & & & & \\
& $\alpha$ & & &           & & & blue    & & & green    & \\
& & & & & & & & & & & \\
\hline
& & & & & & & & & & & \\
& $\beta$  & &  & blue    & & &       & & & red      & \\
& & & & & & & & & & & \\
\hline
& & & & & & & & & & & \\
& $\gamma$ & & &  green    & & & red   & & &          & \\
& & & & & & & & & & & \\
\hline\hline
\end{tabular}
\end{center}
\vskip .5in

\begin{center}
\parbox{4in}
{Table 5. Table of the generation of colors
out of the supercolors.}
\end{center}
\newpage

\begin{center}
\begin{tabular}{ c  c c c c } \hline
& & & & \\
Interaction & Superstrong & Strong & Electromagnetic & Gravity \\
& & & & \\
\hline\hline
& & & & \\
Static  & $\frac{{\mho}_{1}{\mho}_{2}}{4{\pi}r}e
{-{\mu}r}$? & -$\frac{\sqrt{g}\sqrt{g}}{4{\pi}r}e
{-{\mu}r}$  & $\frac{q_{1}q_{2}}{4{\pi}r}e
{-{\mu}r}$  & - $\gamma\frac{m_{1}m_{2}}{r}e {-{\mu}r}$ \\
potential & $\frac{\bar{\hbar}}{{\mu}c}{<}10
{-17}$cm  & $\frac{\hbar}{{\mu}c}{\sim}10
{-13}$cm & ${\mu}=0$ & ${\mu}=0$  \\
& & & & \\
\hline
& & & & \\
Coupling & $\frac{{\mho}_{1}{\mho}_{2}}{4{\pi}\bar{\hbar}c}{>10
{4}}$  & $\frac{g 2}{4{\pi}{\hbar}c}{\approx}10$ & $\frac{e
2}{4{\pi}{\hbar}c}=\frac{1}{137.036}$ & $\frac{{\gamma}{m_{p}}
{2}}{{\hbar}c}=5.76{\times}10 {-36}$ \\
& & & & $m_p$ = proton mass \\
& & & & \\
\hline
& & & & \\
Bosons & $\aleph_{1},\aleph_{2},\aleph_{3}$ & ${\pi} {+},{\pi} {-},{\pi}
{0}$ & photon & graviton \\
& & & & \\
\hline\hline
\end{tabular}
\end{center}
\vskip .6in

\begin{center}
\parbox{5in}
{Table 6. The Six Interactions of Nature. The charge of matter which produces
the superstrong interaction among prequarks is represented by $\mho$. Fermi
constant is given by $\Lambda$. All units are in the CGS system.}
\end{center}

\newpage

\begin{center}
\begin{tabular}{ c c } \hline
& \\
Superweak & Weak \\
& \\
\hline\hline
& \\
$\frac{Q_{1}Q_{2}}{r}e {-{\mu}r}$ & none \\
${\mu}=0$ & \\
& \\
\hline
& \\
${A_{B}, A_{I}} {\approx} 10 {-67}$  & ${\Lambda}{m_p}
{2} = 1.01{\times}10 {-5}$  \\
& \\
& \\
\hline
& \\
symmetron & $W {+},W {-},Z {0}$ \\
&  \\
\hline\hline
\end{tabular}
\end{center}

\newpage
\begin{center}
\begin{tabular}{  c  c l c l  } \hline
& & & &  \\
$n,m,k$ & $E_{C}(Gev)$ & $E_{M}$(Gev) & Error(\%) & $L_{2I.2J}$ \\
& & & &  \\
\hline\hline
0,0,0  & 0.93  & 0.938($N$) & 0.9 & $P_{11}$ \\
\hline
$n+m+k=1$ & 1.24 & 1.232($\Delta$) & 0.6  & $P_{33}$ \\
\hline
$1.24 + \pi$ & 1.38 & 1.44($N$) & 4.3 & $P_{11}$ \\
\hline
$n+m+k=2$ & 1.55 & 1.52($N$) & 1.9 & $D_{13}$ \\
$n+m+k=2$ & 1.55 & 1.535($N$) & 1.0 & $S_{11}$ \\
$n+m+k=2$ & 1.55 & 1.6($\Delta$) & 3.1 & $P_{33}$ \\
$n+m+k=2$ & 1.55 & 1.62($\Delta$) & 4.5 & $S_{31}$ \\
\hline
$1.55 + \pi$ & 1.69 & 1.65($N$) & 2.4 & $S_{11}$ \\
$1.55 + \pi$ & 1.69 & 1.675($N$) & 0.9 & $D_{15}$ \\
$1.55 + \pi$ & 1.69 & 1.68($N$) & 0.6 & $F_{15}$ \\
$1.55 + \pi$ & 1.69 & 1.70($N$) & 0.6 & $D_{13}$ \\
$1.55 + \pi$ & 1.69 & 1.70($\Delta$) & 0.6 & $D_{33}$ \\
$1.55 + \pi$ & 1.69 & 1.71($N$) & 1.2 & $P_{11}$ \\
$1.55 + \pi$ & 1.69 & 1.72($N$) & 1.8 & $P_{13}$ \\
\hline
$n+m+k=3$ & 1.86 & 1.90($N$) & 2.2 & $P_{13}$ \\
$n+m+k=3$ & 1.86 & 1.90($\Delta$) & 2.2 & $S_{31}$ \\
$n+m+k=3$ & 1.86 & 1.905($\Delta$) & 2.4 & $F_{35}$ \\
$n+m+k=3$ & 1.86 & 1.91($\Delta$) & 2.7 & $P_{31}$ \\
$n+m+k=3$ & 1.86 & 1.92($\Delta$) & 3.2 & $P_{33}$ \\
\hline
$1.86 + \pi$ & 2.00 & 1.93($\Delta$) & 3.5 & $D_{35}$ \\
$1.86 + \pi$ & 2.00 & 1.94($\Delta$) & 3.0 & $D_{33}$ \\
$1.86 + \pi$ & 2.00 & 1.95($\Delta$) & 2.5 & $F_{37}$ \\
$1.86 + \pi$ & 2.00 & 1.99($N$) & 0.5 & $F_{17}$ \\
$1.86 + \pi$ & 2.00 & 2.00($N$) & 0 & $F_{15}$ \\
$1.86 + \pi$ & 2.00 & 2.00($\Delta$) & 0 & $F_{35}$ \\
\hline\hline
\end{tabular}
\end{center}
\newpage

\begin{center}
\begin{tabular}{  c  c l c l  } \hline
& & & &  \\
$n,m,k$ & $E_{C}(Gev)$ & $E_{M}$(Gev) & Error(\%) & $L_{2I.2J}$ \\
& & & &  \\
\hline\hline
$n+m+k=4$ & 2.17 & 2.08($N$) & 4.1 & $D_{13}$ \\
$n+m+k=4$ & 2.17 & 2.09($N$) & 3.7 & $S_{11}$ \\
$n+m+k=4$ & 2.17 & 2.10($N$) & 3.2 & $P_{11}$ \\
$n+m+k=4$ & 2.17 & 2.15($\Delta$) & 0.9 & $S_{31}$ \\
$n+m+k=4$ & 2.17 & 2.19($N$) & 0.9 & $G_{17}$ \\
$n+m+k=4$ & 2.17 & 2.20($N$) & 1.4 & $D_{15}$ \\
$n+m+k=4$ & 2.17 & 2.20($\Delta$) & 1.4 & $G_{37}$ \\
$n+m+k=4$ & 2.17 & 2.22($N$) & 2.3 & $H_{19}$ \\
\hline
$2.17 + \pi$ & 2.31 & 2.25($N$) & 2.6 & $G_{19}$ \\
$2.17 + \pi$ & 2.31 & 2.3($\Delta$) & 0.4 & $H_{39}$ \\
$2.17 + \pi$ & 2.31 & 2.35($\Delta$) & 1.7 & $D_{35}$ \\
\hline
$n+m+k=5$ & 2.48 & 2.39($\Delta$) & 3.6 & $F_{37}$ \\
$n+m+k=5$ & 2.48 & 2.40($\Delta$) & 3.2 & $G_{39}$ \\
$n+m+k=5$ & 2.48 & 2.42($\Delta$) & 2.4 & $H_{3,11}$ \\
\hline
$2.48 + \pi$ & 2.62 & 2.60($N$) & 0.8 & $I_{1,11}$ \\
\hline
$n+m+k=6$ & 2.79 & 2.7($N$) & 3.2 & $K_{1,13}$ \\
$n+m+k=6$ & 2.79 & 2.75($\Delta$) & 1.4 & $I_{3,13}$ \\
\hline
$2.79 + \pi$ & 2.93 & 2.95($\Delta$) & 0.7 & $K_{3,15}$ \\
\hline
$n+m+k=7$ & 3.10 & to be found & ? & ? \\
\hline
... & ... & ... & ... & ... \\
\hline\hline
\end{tabular}
\end{center}
\newpage
\begin{center}
\parbox{4.5in}
{Table 7. Baryon states $N$ and $\Delta$. The energies $E_{C}$ were
calculated according to the formula $E_{n,m,k} = 0.31(n+m+k+3) + lm_{\pi}$,
where $l$ is either 0 or 1. $E_{M}$ is the measured energy.  The error
means the absolute value of $(E_{C} - E_{M})/E_{C}$. We are able, of
course, to predict the energies of many other particles.}
\end{center}

\newpage
\pagestyle{empty}
\begin{center}
\begin{tabular}{  c  c l c c  }
\hline\hline
& & & &  \\
State($n,m,k$) & $E_{C}(Gev)$ & $E_{M}$(Gev) & Error(\%) & $L_{I,2J}$ \\
& & & &  \\
\hline\hline
0,0,0  & 1.12  & 1.116($\Lambda$) & 0.4 & $P_{01}$ \\
0,0,0  & 1.12  & 1.193($\Sigma$) & 6.5 & $P_{11}$ \\
\hline
$n+m=1$, k=0 & 1.43 & 1.385($\Sigma$) & 3.2 & $P_{13}$ \\
$n+m=1$, k=0 & 1.43 & 1.405($\Lambda$) & 1.7 & $S_{01}$ \\
$n+m=1$, k=0 & 1.43 & 1.48($\Lambda$) & 3.5 & ? \\
\hline
0,0,1 & 1.62 & 1.52($\Lambda$) & 6.2 & $D_{03}$ \\
0,0,1 & 1.62 & 1.56($\Sigma$) & 3.7 & ? \\
0,0,1 & 1.62 & 1.58($\Sigma$) & 2.5 & $D_{13}$ \\
0,0,1 & 1.62 & 1.60($\Lambda$) & 1.2 & $P_{01}$ \\
0,0,1 & 1.62 & 1.62($\Sigma$) & 0 & $S_{11}$ \\
0,0,1 & 1.62 & 1.66($\Sigma$) & 2.5 & $P_{11}$ \\
0,0,1 & 1.62 & 1.67($\Sigma$) & 3.1 & $D_{13}$ \\
0,0,1 & 1.62 & 1.67($\Lambda$) & 3.1 & $S_{01}$ \\
\hline
$n+m=2$, k=0 & 1.74 & 1.69($\Lambda$) & 2.9 & $D_{03}$ \\
$n+m=2$, k=0 & 1.74 & 1.69($\Sigma$) & 2.9 & ? \\
$n+m=2$, k=0 & 1.74 & 1.75($\Sigma$) & 0.6 & $S_{11}$ \\
$n+m=2$, k=0 & 1.74 & 1.77($\Sigma$) & 1.7 & $P_{11}$ \\
$n+m=2$, k=0 & 1.74 & 1.775($\Sigma$) & 2.0 & $D_{15}$ \\
$n+m=2$, k=0 & 1.74 & 1.80($\Lambda$) & 3.4 & $S_{01}$ \\
$n+m=2$, k=0 & 1.74 & 1.81($\Lambda$) & 4.0 & $P_{01}$ \\
$n+m=2$, k=0 & 1.74 & 1.82($\Lambda$) & 4.6 & $F_{05}$ \\
$n+m=2$, k=0 & 1.74 & 1.83($\Lambda$) & 5.2 & $D_{05}$ \\
\hline
$n+m=1$, k=1 & 1.93 & 1.84($\Sigma$) & 4.7 & $P_{13}$ \\
$n+m=1$, k=1 & 1.93 & 1.88($\Sigma$) & 2.6 & $P_{11}$ \\
$n+m=1$, k=1 & 1.93 & 1.89($\Lambda$) & 2.1 & $P_{03}$ \\
$n+m=1$, k=1 & 1.93 & 1.915($\Sigma$) & 0.8 & $F_{15}$ \\
$n+m=1$, k=1 & 1.93 & 1.94($\Sigma$) & 0.5 & $D_{13}$ \\
\hline\hline
\end{tabular}
\end{center}

\newpage
\begin{center}
\begin{tabular}{  c  c l c l  }
\hline\hline
& & & & \\
State($n,m,k$) & $E_{C}$(GeV) & $E_{M}$(GeV) & Error(\%) & $L_{2I,2J}$ \\
& & & & \\
\hline\hline
$n+m=3$, k=0 & 2.05 & 2.00($\Lambda$) & 2.5 & ? \\
$n+m=3$, k=0 & 2.05 & 2.00($\Sigma$) & 2.4 & $S_{11}$ \\
$n+m=3$, k=0 & 2.05 & 2.02($\Lambda$) & 1.5 & $F_{07}$ \\
$n+m=3$, k=0 & 2.05 & 2.03($\Sigma$) & 1.0 & $F_{17}$ \\
$n+m=3$, k=0 & 2.05 & 2.07($\Sigma$) & 1.0 & $F_{15}$ \\
$n+m=3$, k=0 & 2.05 & 2.08($\Sigma$) & 1.5 & $P_{13}$ \\
\hline
0,0,2 & 2.12 & 2.10($\Sigma$) & 0.9 & $G_{17}$ \\
0,0,2 & 2.12 & 2.10($\Lambda$) & 0.9 & $G_{07}$ \\
0,0,2 & 2.12 & 2.11($\Lambda$) & 0.5 & $F_{05}$ \\
\hline
$m+n=2$, k=1 & 2.24 & 2.25($\Sigma$) & 0.5 & ? \\
\hline
$n+m=4$, k=0 & 2.36 & 2.325($\Lambda$) & 1.5 & $D_{03}$ \\
$n+m=4$, k=0 & 2.36 & 2.35($\Lambda$) & 0.4 & ? \\
\hline
$n+m=1$, k=2 & 2.43 & 2.455($\Lambda$) & 1.0 & ? \\
\hline
$n+m=3$, k=1 & 2.55 & 2.585($\Lambda$) & 1.4 & ? \\
\hline
0,0,3 & 2.62 & 2.62($\Sigma$) & 0 & ? \\
\hline
$n+m=5$, k=0 & 2.67 & to be found & ? & ? \\
\hline
$n+m=2$, k=2 & 2.74 & to be found & ? & ? \\
\hline
$n+m=4$, k=1 & 2.86 & to be found & ? & ? \\
\hline
$n+m=1$, k=3 & 2.93 & to be found & ? & ? \\
\hline
$n+m=6$, k=0 & 2.98 & 3.00($\Sigma$) & 0.7 & ? \\
\hline
$n+m=3$, k=2 & 3.05 & to be found & ? & ? \\
\hline
$n=m=0$, k=4 & 3.12 & to be found & ? & ? \\
\hline
$n+m=5$, k=1 & 3.17 & 3.17($\Sigma$) & 0 & ? \\
\hline
$n+m=2$, k=3 & 3.24 & to be found & ? & ? \\
\hline
... & ... & ... & ...& ... \\
\hline\hline
\end{tabular}
\end{center}
\vskip .3in
\begin{center}
\parbox{4.5in}
{Table 8. Baryon states $\Sigma$ and $\Lambda$. The energies $E_{C}$ were
calculated according to the formula $E_{n,m,k}= 0.31(n+m+2) + 0.5(k+1)$.
$E_{M}$ is the measured energy. The error  means the absolute value of
$(E_{C} - E_{M})/E_{C}$. We are able to predict the energy levels of many
other particles.}
\end{center}

\newpage
\begin{center}
\begin{tabular}{  c  l l l c  } \hline
& & & & \\
State($n,m,k$) & $E_{C}$(Gev) & $E_{M}$(Gev) & Error(\%) & $L_{2I,2J}$ \\
& & & & \\
\hline\hline
0,0,0  & 1.31  & 1.318 & 0.6 & $P_{11}$ \\
\hline
1,0,0  & 1.62 & 1.53 & 5.6 & $P_{13}$ \\
1,0,0 & 1.62 & 1.62 & 0 & ? \\
1,0,0 & 1.62 & 1.69 & 4.3 & ? \\
\hline
n=0, $m+k=1$ & 1.81 & 1.82 & 0.6 & $D_{13}$ \\
\hline
2,0,0 & 1.93 & 1.95 & 1.0 & ? \\
\hline
n=1, $m+k=1$ & 2.12 & 2.03 & 4.2 & ? \\
n=1, $m+k=1$ & 2.12 & 2.12 & 0 & ? \\
\hline
n=3, $m=k=0$ & 2.24 & 2.25 & 0.5 & ? \\
\hline
n=0, $m+k=2$ & 2.31 & 2.37 & 2.6 & ? \\
\hline
n=2, $m+k=1$ & 2.43 & to be found & ? & ? \\
\hline
n=4, $m=k=0$ & 2.55 & 2.5 & 2.0 & ? \\
\hline
n=1, $m+k=2$ & 2.62 & to be found & ? & ? \\
... & ... & ... & ... & ... \\
\hline\hline
\end{tabular}
\end{center}
\vskip .3in

\begin{center}
\parbox{4.5in}
{Table 9. Baryon states $\Xi$. The energies $E_{C}$ were
calculated according to the formula $E_{n,m,k}= 0.31(n+1) + 0.5(m+k+2)$.
$E_{M}$ is the measured energy. The error means the absolute
value of $(E_{C} - E_{M})/E_{C}$. We are able, of course, to
predict the energies of many other particles.}
\end{center}

\newpage\begin{center}
\begin{tabular}{  c  l c l  } \hline
& & &  \\
State($n,m,k$) & $E_{C}$(Gev) & $E_{M}$(Gev) & Error(\%) \\
& & &  \\
\hline\hline
0,0,0  & 1.5  & 1.675 & 11.7 \\
\hline
$n+m+k=1$ & 2.0 & 2.25 & 12.5 \\
\hline
$2.0 + \pi$ & 2.14 & 2.38 & 11.2 \\
\hline
$n+m+k=2$ & 2.5 & 2.47 & 1.2 \\
\hline
$2.5 + \pi$ & 2.64 & to be found & ? \\
\hline
$n+m+k=3$ & 3.0 & to be found & ? \\
\hline
... & ... & ... & ... \\
\hline\hline
\end{tabular}
\end{center}
\vskip .3in

\begin{center}
\parbox{4.5in}
{Table 10. Baryon states $\Omega$. The energies $E_{C}$ were
calculated according to the formula $E_{n,m,k}= 0.5(n+m+k+3)$.
$E_{M}$ is the measured energy. The error means the absolute
value of $(E_{C} - E_{M})/E_{C}$. We are able, of course, to
predict the energies of many other particles.}
\end{center}

\newpage\begin{center}
\begin{tabular}{  c  l c l  } \hline
& & &  \\
State($n,m,k$) & $E_{C}$(Gev) & $E_{M}$(Gev) & Error(\%) \\
& & &  \\
\hline\hline
0,0,0 & 2.32 & 2.285($\Lambda_{c}$) & 1.5 \\
0,0,0 & 2.32 & 2.455($\Sigma_{c}$) & 5.8 \\
\hline
$n+m=1$, k=0 & 2.63 & to be found & ? \\
\hline
$n+m=2$, k=0 & 2.94 & to be found & ? \\
\hline
... & ... & ... & ...\\
\hline\hline
\end{tabular}
\end{center}
\vskip .6in

\begin{center}
\parbox{4.5in}
{Table 11. Baryon states $\Lambda_{c}$ and $\Sigma_{c}$. The
energies $E_{C}$ were calculated according to the formula
$E_{n,m,k}= 0.31(n+m+2) + 1.7(k+1)$. $E_{M}$ is the measured
energy. The error means the absolute value of $(E_{C} - E_{M})/E_{C}$.
We are able, of course, to predict the energies of many other
particles.}
\end{center}

\newpage\begin{center}
\begin{tabular}{  c  l c l  } \hline
& & &  \\
State($n,m,k$) & $E_{C}$(Gev) & $E_{M}$(Gev) & Error(\%) \\
& & &  \\
\hline\hline
0,0,0  & 2.51  & 2.47 & 1.6 \\
\hline
1,0,0 & 2.82 & to be found & ? \\
\hline
0,1,0 & 3.01 & to be found & ? \\
\hline
... & ... & ... & ...\\
\hline\hline
\end{tabular}
\end{center}
\vskip .3in
\begin{center}
\parbox{4.5in}
{Table 12. Baryon states $\Xi_{c}$. The energies $E_{C}$ were
calculated according to the formula $E_{n,m,k}= 0.31(n+1) + 0.5(m+1)
+ 1.7(k+1)$. $E_{M}$ is the measured energy. The error means the absolute
value of $(E_{C} - E_{M})/E_{C}$. We are able, of course, to
predict the energies of many other particles.}
\end{center}

\newpage\begin{center}
\begin{tabular}{  c  l c l  } \hline
& & &  \\
State($n,m,k$) & $E_{C}$(Gev) & $E_{M}$(Gev) & Error(\%) \\
& & &  \\
\hline\hline
0,0,0  & 2.7  & to be found & ? \\
\hline
$n+m=1$, k=0 & 3.2 & to be found & ? \\
\hline
$n+m=2$, k=0 & 3.7 & to be found & ? \\
\hline
... & ... & ... & ...\\
\hline\hline
\end{tabular}
\end{center}
\vskip .3in
\begin{center}
\parbox{4.5in}
{Table 13. Baryon states $\Omega_{c}$. The energies $E_{C}$ were
calculated according to the formula $E_{n,m,k}= 0.5(n+m+2) + 1.7(k+1)$.
$E_{M}$ is the measured energy. The error means the absolute
value of $(E_{C} - E_{M})/E_{C}$. We are able, of course, to
predict the energies of many other particles.}
\end{center}

\newpage
\begin{center}
\begin{tabular}{  c  c c c c  } \hline
& & & & \\
State($n$) & $E_{u}=E_{d}$(GeV) & $E_{s}$(GeV) & $E_{c}$(GeV) & $E_{b}$(GeV) \\
& & & & \\
\hline\hline
0 & 0.31 & 0.5 & 1.7 & 5 \\
\hline
1 & 0.93 & 1.5 & 5.1 & 15 \\
\hline
2 & 1.55 & 2.5 & 8.5 & 25 \\
\hline
3 & 2.17 & 3.5 & 11.9 & 35 \\
\hline
...& ... & ... & ... & ... \\
\hline\hline
\end{tabular}
\end{center}
\vskip .3in
\begin{center}
\parbox{4.5in}
{Table 14. The energy states of quarks u, d, c, s and b.}
\end{center}

\newpage\begin{center}
\begin{tabular}{  c  c l  } \hline
& & \\
State & $E_{n}$(MeV) & Particles \\
& & \\
\hline\hline
$\pi_{0}$ & 135-140 & $\pi {\pm}(140)$, $\pi {0}(135)$ \\
\hline
$\pi_{1}$ & 405-420 & ? \\
\hline
$\pi_{2}$ & 675-700 & $\epsilon({\sim}700)$ \\
\hline
$\pi_{3}$ & 945-980 & ${\eta}'(958)$, $f_{0}(974)$, $a_{0}(983)$ \\
\hline
$\pi_{4}$ & 1215-1260 & $h_{1}(1170)$, $b_{1}(1232)$, $a_{1}(1260)$, \\
& & $f_{2}(1275)$, $f_{1}(1282)$, $\eta(1295)$, \\
& & $\pi(1300)$ \\
\hline
$\pi_{5}$ & 1485-1540 & $\rho(1465)$, $f_{1}(1512)$, $f_{2}'(1525)$ \\
\hline
$\pi_{6}$ & 1755-1820 & $\pi(1770)$, $\phi_{3}(1854)$ \\
\hline
$\pi_{7}$ & 2025-2100 & $f_{2}(2011)$, $f_{4}(2049)$ \\
\hline
$\pi_{8}$ & 2295-2380 & $f_{2}(2297)$, $f_{2}(2339)$ \\
\hline
$\pi_{9}$ & 2565-2660 & to be found \\
\hline
$\pi_{10}$ & 2835-2940 & to be found \\
\hline
... & ... & ... \\
\hline
\hline
\end{tabular}
\end{center}
\vskip .3in
\begin{center}
\parbox{6in}
{Table 15. The energy states of mesons of the pion family, that is, pion
oscillations. $E_{n}$ was calculated by the formula ${\hbar}\nu(n + 0.5)$
taking for $\hbar\nu$ the values 270MeV and 280MeV. The error between
experimental and calculated values is in general below 3\%, for every
particle.}
\end{center}

\newpage\begin{center}
\begin{tabular}{  c  c l  } \hline
& & \\
State & $E_{n}$(MeV) & Particles \\
& & \\
\hline\hline
$\pi_{0 + 1}$ & 540-560 & $\eta(548)$ \\
\hline
$\pi_{0 + 2}$ & 810-840 & $\rho(768)$, $\omega(782)$ \\
\hline
$\pi_{0 + 3}$ & 1080-1120 & $\phi(1019)$ \\
\hline
$\pi_{0 + 4}$ & 1350-1400 & $a_{2}(1318)$, $\omega(1394)$, $f_{0}(1400)$,
$f_{1}(1426)$, $\eta(1420)$ \\
\hline
$\pi_{0 + 5}$ & 1620-1680 & $f_{0}(1587)$, $\omega(1594)$, $\omega_{3}(1668)$,
$\pi_{2}(1670)$, \\
& & $\phi(1680)$, $\rho_{3}(1690)$, $\rho(1700)$, $f_{0}(1709)$ \\
\hline
$\pi_{0 + 6}$ & 1890-1960 & $\phi_{3}(1854)$ \\
\hline
$\pi_{0 + 7}$ & 2160-2240 & to be found \\
\hline
... & ... & ... \\
\hline
\hline
\end{tabular}
\end{center}
\vskip .3in
\begin{center}
\parbox{6.5in}
{Table 16. The energy states of mesons of the pion family which are
the result of overtones of the $\pi_{0}$ with all states $\pi_{n}$.
The error between experimental and calculated values is in general
below 5\%, for every particle.}
\end{center}

\newpage\begin{center}
\begin{tabular}{  c  c l  } \hline
& & \\
State($K_{n}$) & $E_{n}$(MeV) & Particles \\
& & \\
\hline\hline
$K_{0}$ & 498 & $K {\pm}(494)$, $K {0}(498)$ \\
\hline
$K_{1}$ & 1494 & $f_{2}'(1525)$, $f_{1}(1512)$, $\rho(1465)$ \\
\hline
$K_{2}$ & 2490 & to be found \\
\hline
... & ... & ... \\
\hline\hline
\end{tabular}
\end{center}
\vskip .3in
\begin{center}
\parbox{6in}
{Table 17. The energy states of kaons. $E_{n}$ was calculated by the formula
${\hbar}\nu(n + 0.5)$ taking for $\hbar\nu$ the value of 996MeV. The error
between experimental and calculated values is in general below 2\%, for every
particle.}
\end{center}

\newpage\begin{center}
\begin{tabular}{  c  c l  } \hline
& & \\
State & $E_{n}$(MeV) & Particles \\
& & \\
\hline\hline
$K_{0} + \pi_{0}$ & 633-638 & ? \\
\hline
$K_{0} + \pi_{1}$ & 903-918 & $K {*(\pm)}(892)$, $K {*(0)}(896)$ \\
\hline
$K_{0} + \pi_{0 + 1}$ & 1038-1058 & $f_{0}(974)$, $a_{0}(983)$, $\phi(1019)$ \\
\hline
$K_{0} + \pi_{2}$ & 1173-1198 & $b_{1}(1232)$, $h_{1}(1170)?$ \\
\hline
$K_{0} + \pi_{0 + 2}$ & 1308-1338 & $f_{2}(1270)$, $f_{1}(1282)$,
$a_{2}(1318)$, $K_{1}(1270)$, \\
& & $a_{1}(1260)?$, $\eta(1295)?$, $\pi(1300)?$ \\
\hline
$K_{0} + \pi_{3}$ & 1443-1478 & $f_{0}(1400)$, $f_{1}(1426)$, $\eta(1440)$,
$K_{1}(1402)$, \\
& & $K {*}(1412)$, $K {*}_{0}(1429)$, $K {*(\pm)}_{2}(1425)$,
$K {*(0)}_{2}(1432)$ \\
\hline
$K_{0} + \pi_{0 + 3}$ & 1578-1618 & $\omega(1594)?$ \\
\hline
$K_{0} + \pi_{4}$ & 1713-1758 & $\pi_{2}(1670)$, $\phi(1680)$,
$\rho_{3}(1691)$, $\rho(1700)$, \\
& & $f_{0}(1709)$, $K {*}(1714)$, $K_{2}(1768)$, $K_{3} {*}(1770)$ \\
\hline
$K_{0} + \pi_{0 + 4}$ & 1848-1898 & $\phi_{3}(1854)$ \\
\hline
$K_{0} + \pi_{5}$ & 1983-2038 & $f_{4}(2049)$, $K
{*}_{4}(2045)$, $f_{2}(2011)?$ \\
\hline
$K_{0} + \pi_{0 + 5}$ & 2118-2178 & to be found \\
\hline
$K_{0} + \pi_{6}$ & 2253-2318 & $f_{2}(2297)?$, $f_{2}(2339)?$  \\
\hline
$K_{0} + \pi_{0 + 6}$ & 2388-2458 & to be found \\
\hline
... & ... & ... \\
\hline\hline
\end{tabular}
\end{center}
\vskip .3in
\begin{center}
\parbox{4.5in}
{Table 18. The energies of mesons which are the result of the
overtones of $K_{0}$ with pions. The interrogation mark means that the
decay of the particle into kaons has not yet been found experimentally.
The error between experimental and calculated values is below 3\%, in
general.}\

\newpage\begin{center}
\begin{tabular}{  c  c l  } \hline
& & \\
State & $E_{n}$(MeV) & Particles \\
& & \\
\hline\hline
$D_{0}$ & 1864 & $D {\pm}(1869)$, $D {0}(1864)$ \\
\hline
$D_{1}$ & 5595 & to be found \\
\hline
$D_{2}$ & 9325 & to be found \\
\hline
... & ... & ... \\
\hline\hline
\end{tabular}
\end{center}
\vskip .6in
\begin{center}
\parbox{4.5in}
{Table 19. The energy states of $D$ mesons. $E_{n}$ was calculated
according to the formula $E_{n} = \hbar\nu(n + 0.5)$ with
$\hbar\nu = 3730$MeV.}\end{center}

\newpage\begin{center}
\begin{tabular}{  c  c l  } \hline
& & \\
State & $E_{n}$(MeV) & Particles \\
& & \\
\hline\hline
$D_{0} + \pi_{0}$ & 2005 & $D {\ast\pm}(2010)$, $D {\ast{0}}(2007)$ \\
\hline
$D_{0} + \pi_{1}$ & 2284 & to be found \\
\hline
$D_{0} + \pi_{0 + 1}$ & 2424 & $D_{1} {0}(2424)$, $D_{2}
{\ast}(2459)$ \\
\hline
$D_{0} + \pi_{2}$ & 2564 & to be found \\
\hline
... & ... & ... \\
\hline\hline
\end{tabular}
\end{center}
\vskip .3in
\begin{center}
\parbox{4.5in}
{Table 20. The energy states of $D$ mesons which are the result of
overtones of $D_{0}$ with $\pi_{n}$. The error between experimental and
calculated values is in general below 1.5\%, for every particle.}
\end{center}

\newpage\begin{center}
\begin{tabular}{  c  c l  } \hline
& & \\
State & $E_{n}$(MeV) & Particles \\
& & \\
\hline\hline
$(D_{S})_{0}$ & 1969 & $D_{S} {\pm}(1969)$ \\
\hline
$(D_{S})_{1}$ & 5907 & to be found \\
\hline
$(D_{S})_{2}$ & 9845 & to be found \\
\hline
... & ... & ... \\
\hline\hline
\end{tabular}
\end{center}
\vskip .3in
\begin{center}
\parbox{4.5in}
{Table 21. The energy states of $D_{S}$ mesons. $E_{n}$ was calculated
according to the formula $E_{n} = \hbar\nu(n + 0.5)$ with
$\hbar\nu = 3938$MeV.}
\end{center}

\newpage\begin{center}
\begin{tabular}{  c  c l  } \hline
& & \\
State & $E_{n}$(MeV) & Particles \\
& & \\
\hline\hline
$(D_{S})_{0} + \pi_{0}$ & 2109 & $D_{S} {\ast\pm}(2110)$ \\
\hline
$(D_{S})_{0} + \pi_{1}$ & 2389 & to be found \\
\hline
$(D_{S})_{0} + \pi_{0 + 1}$ & 2529 & $D_{S1} {\pm}(2537)$ \\
\hline
$(D_{S})_{0} + \pi_{2}$ & 2669 & to be found \\
\hline
$(D_{S})_{0} + \pi_{0 + 2}$ & 2809 & to be found \\
... & ... & ... \\
\hline\hline
\end{tabular}
\end{center}
\vskip .3in
\begin{center}
\parbox{4.5in}
{Table 22. The energy states of $D_{S}$ mesons which are the result of
overtones of $(D_{S})_{0}$ with $\pi_{n}$. The error between experimental and
calculated values is below 0.5\%, for every particle.}
\end{center}

\newpage\begin{center}
\begin{tabular}{  c  c l  } \hline
& & \\
State & $E_{n}$(MeV) & Particles \\
& & \\
\hline\hline
$B_{0}$ & 5279 & $B {0}(5279)$, $B {\pm}(5279)$ \\
\hline
$B_{1}$ & 15837 & to be found \\
\hline
$B_{2}$ & 26395 & to be found \\
\hline
... & ... & ... \\
\hline\hline
\end{tabular}
\end{center}
\vskip .3in
\begin{center}
\parbox{4.5in}
{Table 23. The energy states of $B$ mesons. $E_{n}$ was calculated
according to the formula $E_{n} = \hbar\nu(n + 0.5)$ with
$\hbar\nu = 10558$MeV.}\end{center}

\newpage\begin{center}
\begin{tabular}{  c  c l  } \hline
& & \\
State & $E_{n}$(MeV) & Particles \\
& & \\
\hline\hline
$B_{0} + \pi_{0}$ & 5414-5419 & to be found \\
\hline
$B_{0} + \pi_{1}$ & 5684-5699 & to be found \\
\hline
... & ... & ... \\
\hline\hline
\end{tabular}
\end{center}
\vskip .4in
\begin{center}
\parbox{4.5in}
{Table 24. The energy states of $B$ mesons which are the result of
overtones of $B_{0}$ with $\pi_{n}$.}\end{center}

\newpage\begin{center}
\begin{tabular}{  c  c l  } \hline
& & \\
State & $E_{n}$(MeV) & Particles \\
& & \\
\hline\hline
$c\bar{c}_{0}$ & 2978 & $\eta_{c}(2978)$ \\
\hline
$c\bar{c}_{1}$ & 8967 & to be found \\
\hline
$c\bar{c}_{2}$ & 14890 & to be found \\
\hline
... & ... & ... \\
\hline\hline
\end{tabular}
\end{center}
\vskip .3in
\begin{center}
\parbox{4.5in}
{Table 25. The energy states of $c\bar{c}$ mesons. $E_{n}$ was calculated
according to the formula $E_{n} = \hbar\nu(n + 0.5)$ with
$\hbar\nu = 5956$MeV.}\end{center}

\newpage\begin{center}
\begin{tabular}{  c  c l  } \hline
& & \\
State & $E_{n}$(MeV) & Particles \\
& & \\
\hline\hline
$c\bar{c}_{0} + \pi_{0}$ & 3113-3118 & $J/{\Psi}(3097)$ \\
\hline
$c\bar{c}_{0} + \pi_{1}$ & 3383-3398 & $\chi_{c0}(3415)$ \\
\hline
$c\bar{c}_{0} + \pi_{0 + 1}$ & 3518-3538 & $\chi_{c1}(3510)$, $\chi_{c2}(3556)$
\\ \hline
$c\bar{c}_{0} + \pi_{2}$ & 3653-3668 & $\Psi(3686)$ \\
\hline
$c\bar{c}_{0} + \pi_{0 + 2}$ & 3788-3818 & $\Psi(3770)$ \\
\hline
$c\bar{c}_{0} + \pi_{3}$ & 3923-3958 & ? \\
\hline
$c\bar{c}_{0} + \pi_{0 + 3}$ & 4058-4098 & $\Psi(4040)$ \\
\hline
$c\bar{c}_{0} + \pi_{4}$ & 4193-4238 & $\Psi(4159)$ \\
\hline
$c\bar{c}_{0} + \pi_{0 + 4}$ & 4328-4378 & $\Psi(4415)$ \\
\hline
$c\bar{c}_{0} + \pi_{5}$ & 4463-4518 & to be found \\
\hline
$c\bar{c}_{0} + \pi_{0 + 5}$ & 4598-4658 & to be found \\
... & ... & ... \\
\hline\hline
\end{tabular}
\end{center}
\vskip .3in
\begin{center}
\parbox{4.5in}
{Table 26. The energy states of $c\bar{c}$ mesons which are the result of
overtones of $c\bar{c}_{0}$ with $\pi_{n}$. The error between experimental and
calculated values is below 1\%, for every particle.}\end{center}

\newpage\begin{center}
\begin{tabular}{  c  c l  } \hline
& & \\
State & $E_{n}$(MeV) & Particles \\
& & \\
\hline\hline
$b\bar{b}_{0}$ & 9460 & $\Upsilon(9460)$ \\
\hline
$b\bar{b}_{1}$ & 28380 & to be found \\
\hline
$b\bar{b}_{2}$ & 47300 & to be found \\
\hline
... & ... & ... \\
\hline\hline
\end{tabular}
\end{center}
\vskip .3in
\begin{center}
\parbox{4.5in}
{Table 27. The energy states of $b\bar{b}$ mesons. $E_{n}$ was calculated
according to the formula $E_{n} = \hbar\nu(n + 0.5)$ with
$\hbar\nu = 18920$MeV.}\end{center}

\newpage\begin{center}
\begin{tabular}{  c  c l  } \hline
& & \\
State & $E_{n}$(MeV) & Particles \\
& & \\
\hline\hline
$b\bar{b}_{0} + \pi_{0}$ & 9595-9600 & ? \\
\hline
$b\bar{b}_{0} + \pi_{1}$ & 9865-9880 & $\chi_{b0}(9860)$, $\chi_{b1}(9892)$,
 \\
\hline
$b\bar{b}_{0} + \pi_{0 + 1}$ & 10000-10020 & $\Upsilon(10023)$ \\
\hline
$b\bar{b}_{0} + \pi_{2}$ & 10135-10160 & ? \\
\hline
$b\bar{b}_{0} + \pi_{0 + 2}$ & 10270-10300 & $\chi_{b0}(10235)$,
$\chi_{b1}(1025)$ \\
\hline
$b\bar{b}_{0} + \pi_{3}$ & 10405-10440 & ? \\
\hline
$b\bar{b}_{0} + \pi_{0 + 3}$ & 10540-10580 & $\Upsilon(10580)$ \\
\hline
$b\bar{b}_{0} + \pi_{4}$ & 10675-10710 & ? \\
\hline
$b\bar{b}_{0} + \pi_{0 + 4}$ & 10810-10860 & $\Upsilon(10860)$ \\
\hline
$b\bar{b}_{0} + \pi_{5}$ & 10945-11000 & $\Upsilon(11020)$ \\
\hline
$b\bar{b}_{0} + \pi_{0 + 5}$ & 11080-11140 & to be found \\
\hline
$b\bar{b}_{0} + \pi_{6}$ & 11215-11280 & to be found \\
... & ... & ... \\
\hline\hline
\end{tabular}
\end{center}
\vskip .3in
\begin{center}
\parbox{4.5in}
{Table 28. The energy states of $b\bar{b}$ mesons which are the result of
overtones of $b\bar{b}_{0}$ with $\pi_{n}$. The error between experimental and
calculated values is below 1\%, for every particle.}
\end{center}

\newpage\vskip .3in
\large
\noindent
Complete address:
\vskip .3in
\noindent
M\'{a}rio Everaldo de Souza,
\newline
\noindent
Universidade Federal de Sergipe
\newline
\noindent
Departamento de F\'{\i}sica - CCET,
\newline
\noindent
49000 Aracaju, Sergipe, Brazil
\newline
\noindent
Phone no. (55)(79)241-2848, extension 354
\newline
\noindent
Fax no. (55)(79)241-3995
\newline
\noindent
e-mail DFIMES@BRUFSE.BITNET
\end{center}
\end{document}